\documentclass{egpubl}
\usepackage{eg2025}
\usepackage{amssymb}
\usepackage{amsmath}

 \usepackage{xcolor}
\usepackage{soul} 
\definecolor{darkgreen}{rgb}{0.0, 0.5, 0.0} 

\newcommand{\removed}[1]{}
\newcommand{\added}[1]{#1}

\newcommand{\approvedremoved}[1]{}               

\ConferencePaper        

\CGFccby

\usepackage[T1]{fontenc}
\usepackage{dfadobe}  
\usepackage{cite}  
\usepackage{preamble}
\BibtexOrBiblatex

\electronicVersion
\PrintedOrElectronic

\title[Spherical Neural Surfaces]
      {Neural Geometry Processing via Spherical Neural Surfaces}

\author[Williamson and Mitra]
{\parbox{\textwidth}{\centering Romy Williamson$^{1}$\orcid{0009-0000-9833-7821}
and Niloy J. Mitra$^{1,2}$\orcid{0000-0002-2597-0914} 
        }
        \\
{\parbox{\textwidth}{\centering $^1$University College London \;\;\;
         $^2$Adobe Research
       }
}
}

\begin{document}

\teaser{ 
\centering
\includegraphics[width=\textwidth]{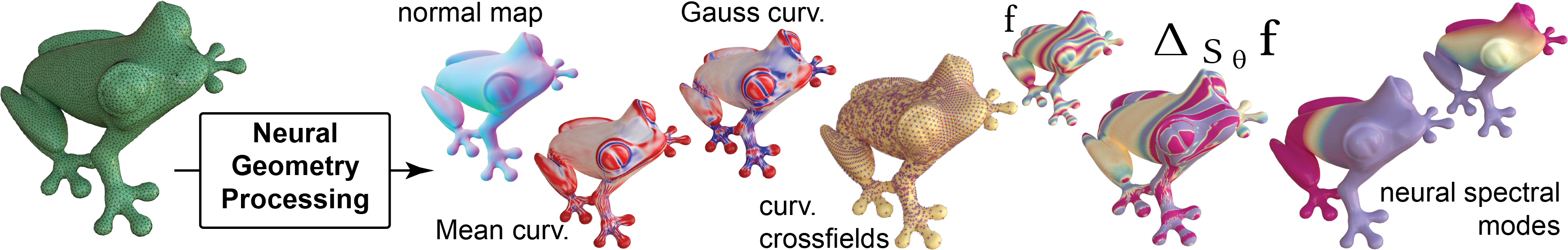}
\mycaption{Neural Geometry Processing}{We encode input genus-0 surfaces as overfitted neural networks and propose operators on them. Specifically, we describe how to compute the normals and the First and Second Fundamental Forms, and hence compute curvatures. We also define a Laplace-Beltrami operator directly using the neural representation, thus enabling processing of scalar (or vector) fields on the underlying surface, and we find the smallest eigenmodes of the Laplace-Beltrami operator via a novel optimization scheme. We avoid unnecessary discretization in the estimates, as commonly encountered while using a traditional surface representation (e.g., a polygonal mesh). }
\label{fig:teaser}
}


\maketitle

\begin{abstract}

%
Neural surfaces (e.g., neural map encoding, deep implicit, and neural radiance fields) have recently gained popularity because of their generic structure (e.g., multi-layer perceptron) and easy integration with modern learning-based setups.
Traditionally, we have a rich toolbox of geometry processing algorithms designed for polygonal meshes to analyze and operate on surface geometry. 
Without an analogous toolbox, neural representations are typically discretized and converted into a mesh, before applying any geometry processing algorithm. This is unsatisfactory and, as we demonstrate, unnecessary. 
In this work, we propose a \textit{spherical neural surface} representation for genus-0 surfaces and demonstrate how to compute core geometric operators directly on this representation. Namely, we estimate surface normals and first and second fundamental forms of the surface, as well as compute surface gradient, surface divergence and Laplace Beltrami operator on scalar/vector fields defined on the surface. Our representation is fully seamless, overcoming a key limitation of similar explicit representations such as Neural Surface Maps~\cite{morreale2021neural}.  
These operators, in turn, enable geometry processing directly on the neural representations \textit{without} any unnecessary meshing. We demonstrate illustrative applications in (neural) spectral analysis, heat flow and mean curvature flow, and evaluate robustness to isometric shape variations. We propose theoretical formulations and validate their numerical estimates, against analytical estimates, mesh-based baselines, and neural alternatives, where available. By systematically linking neural surface representations with classical geometry processing algorithms, we believe this work can become a key ingredient in enabling \textit{neural geometry processing}. Code is available via the \href{https://geometry.cs.ucl.ac.uk/projects/2025/sns/}{project webpage}.

\end{abstract}

\section{Introduction}


The use of neural networks to represent surfaces has gained rapid popularity as a generic surface representation. Many variants of this formulation~\cite{deepSDF:19,occupancyNet:18,groueix2018,qi2016pointnet} have been proposed, such as directly representing a single surface as an overfitted network, or a collection of surfaces, as encoder-decoder networks. The `simplicity' of the approach lies in the network architectures (e.g., MLPs, UNets) being universal function approximators. Such deep representations can either be overfitted to given meshes~\cite{DBLP:journals/corr/abs-2009-09808,morreale2021neural} or directly optimized from a set of posed images (i.e., images with camera information), as popularized by the NeRF formulation~\cite{nerf:21}. Although there are some isolated and specialized attempts~\cite{novell_implicitNeuralSurface_22,chetan2023accurate,yang2021geometry}, no established set of operators directly works on a neural object, encoding an \textit{explicit} surface, to facilitate easy integration with existing geometry processing tasks.

Traditionally, triangle meshes, more generally polygonal meshes, are used to represent surfaces. They are simple to define, easy to store, and benefit from having a wide range of available algorithms supporting them. Most geometry processing algorithms rely on the following core operations: (i)~estimating differential quantities (i.e., surface normals, and first and second fundamental forms) at any surface location; 
(ii)~computing surface gradients and surface divergence on scalar/vector fields; 
and (iii)~defining a Laplace Beltrami operator (i.e., the generalization of the Laplace operator on a curved surface) for analyzing the shape itself or functions on the shape. 

Meshes, however, are discretized surface representations, and accurately carrying over definitions from continuous to discretized surfaces is surprisingly difficult. For example, the area of discrete differential geometry~(DDG)~\cite{ddg:06:course}
was entirely invented to meaningfully carry over continuous differential geometry concepts to the discrete setting. However, many commonly used estimates vary with the mesh quality (i.e., distribution of vertices and their connections), even when encoding the same underlying surface. In the case of the Laplace-Beltrami operator, the `no free lunch' result~\cite{noFreeLunch:07} tells us that we cannot hope for any particular discretization of the Laplace-Beltrami operator to simultaneously satisfy all of the properties of its continuous counterpart. In practice, the most suitable choice of discretization, such as triangular or, more generally, polygonal meshes, can change depending on the specifics of the target application (see~\cite{LBcourse:23} for a recent survey). 

In the absence of native geometric operators for neural representations, the common workaround for applying a geometry processing algorithm to neural representations is discretizing these representations into an explicit mesh (e.g., using Marching cubes) and then relying on traditional mesh-based operators. Unfortunately, this approach inherits the disadvantages of having an explicit mesh representation, as discussed above. 
A similar motivation prompted researchers to pursue a grid-free approach for volumetric analysis using Monte Carlo estimation for geometry processing~\cite{monteCarlo:20} or neural field convolution~\cite{Nsampi2023NeuralFC} to perform general continuous convolutions with continuous signals such as neural fields.

We propose a different approach to discretization-free geometry processing. First, we create a smooth and differentiable representation --- \textit{spherical neural surfaces~(SNS)} --- where we encode individual genus-0 surfaces using dedicated neural networks. Having a differentiable encoding allows one to directly compute surface normals, tangent planes, and Fundamental Forms directly with continuous differential geometry.
The First Fundamental Form allows us to compute attributes of the parametrization, such as local area distortion. 
The Second Fundamental form allows us to compute surface attributes such as mean curvature, Gauss curvature, and principal curvature directions. Essentially, encoding surfaces as seamless neural maps unlocks definitions from differential geometry.

Other estimates require more thought. We implement two methods to apply the \textit{continuous} Laplace Beltrami operator to any differentiable scalar or vector field (using Equations~\ref{eqn:LBdefn} and \ref{eqn:LB_meancurv}). Additionally, to enable spectral analysis on the resultant Laplace Beltrami operator, we propose a novel optimization scheme via a neural representation of scalar fields that allows us to extract the lowest $k$ spectral modes, using the variational formulation of the Dirichlet eigenvalue problem. We use Monte Carlo estimates to approximate the continuous inner products and integration of functions on the (neural) surface. Our framework of spherical neural surfaces makes optimizing many expressions involving surface gradients and surface integrals of smooth functions possible. We hope that our setup opens the door to solving other variational problems posed on smooth (neural) surfaces.

\begin{figure}[b!]
    \centering
    \includegraphics[width=\columnwidth]{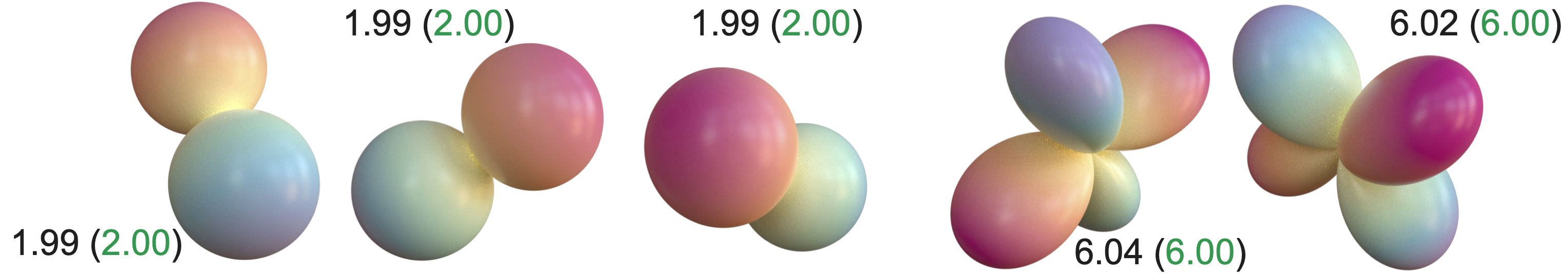}
    \mycaption{Neural Spectral Basis against Analytical Solutions}
     {The sphere is a genus-0 surface for which the eigenspaces of the Laplace-Beltrami operator are very well understood: the eigenspaces are the spaces spanned by linear combinations of the spherical harmonics for each frequency. This means that each of our optimized eigenfunctions should belong to one of these eigenspaces, and therefore, the Rayleigh quotient should match the energy level of the corresponding spherical harmonics. For evaluation, we compute the Rayleigh quotient using a large number of samples for Monte Carlo integration, since we cannot analytically compute the integral. For reference, we present the ground truth Rayleigh quotients, in green. 
    }
\label{fig:LB_on_sphere_analytical}
\vspace*{-.1in}
\end{figure}

\begin{figure}[t!]
    \centering
    \includegraphics[width=\columnwidth]{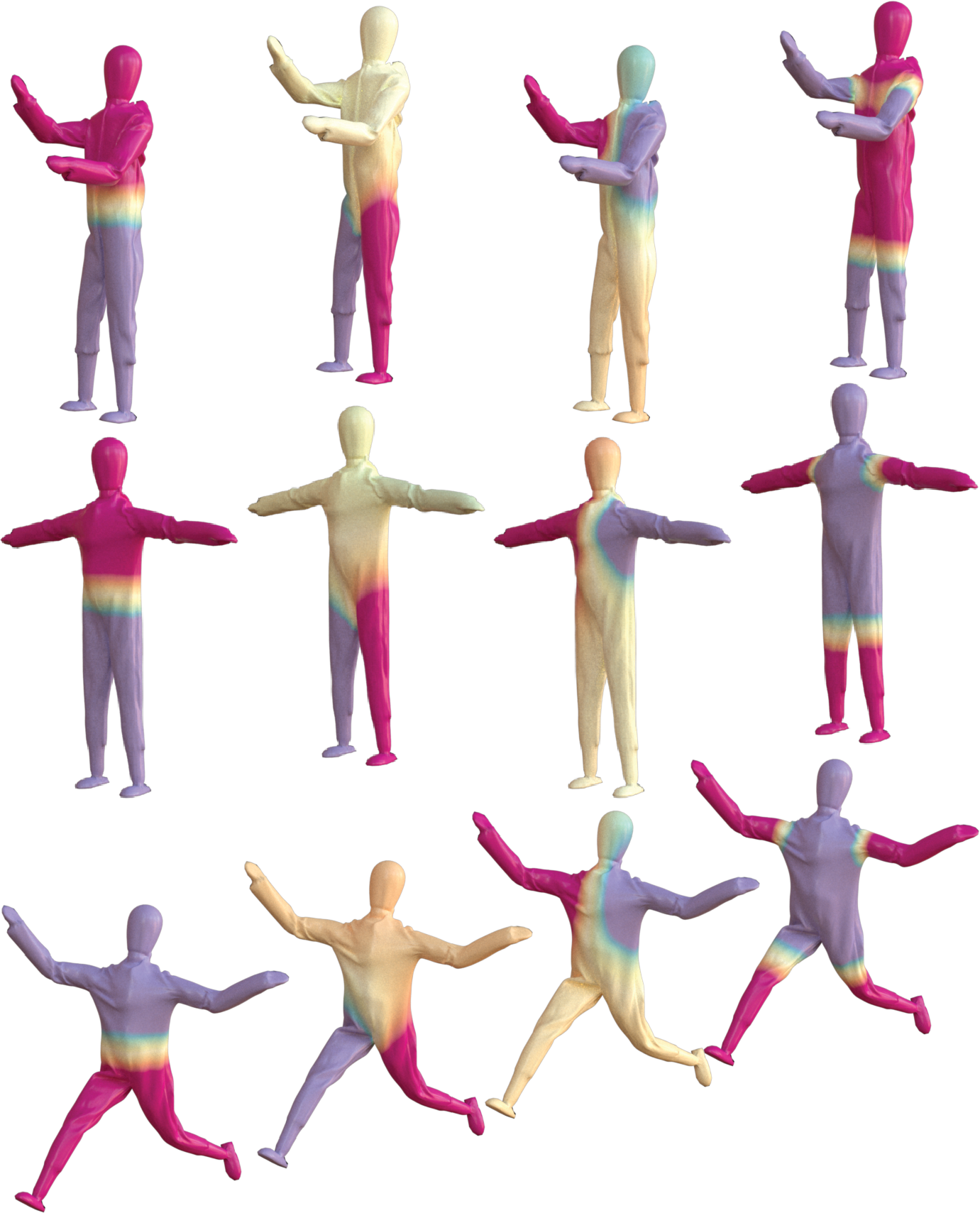}
    \mycaption{Spectral SNS Modes across Isometric Poses}
    {
    If we compute the first few eigenfunctions on SNSs that represent near-isometric surfaces, and order the functions according to their Rayleigh quotients, we see a clear correspondence between corresponding eigenfunctions on each surface. Although our computation of the LBO and its spectrum involves extrinsic terms such as mean curvature and normals, the Laplace Beltrami operator is intrinsic so its spectrum does not depend on the particular embedding of a surface into $\mathbb{R}^3$, therefore the eigenmodes of isometric SNSs should appear similar (up to a possible sign change). 
    }
\label{fig:neural_sparse_matching}
\end{figure}

We evaluate our proposed framework on standard geometry processing tasks. In the context of mesh input, we assess the robustness of our estimates to different meshing of the same underlying surfaces (Figure \ref{fig:LapBelt_robustness_meshing}) and isometrically related shapes (Figure \ref{fig:neural_sparse_matching}). 
%
We compare the results against ground truth values when they are computable (e.g., analytical functions -- see Figure \ref{fig:differentialEstimates_analytical} and Figure \ref{fig:LB_on_sphere_analytical}) and contrast them against alternatives. Specifically, we compare against traditional mesh-based methods (e.g., curvature estimates, cotan LBO, -- see Figures~\ref{fig:scalarLB}, \ref{fig:LapBelt_robustness_meshing}, \ref{fig:LapBelt_robustness_meshing_igea}) or neural alternatives (e.g., NGP~\cite{yang2021geometry} and neural implicits~\cite{novell_implicitNeuralSurface_22} -- see Figure~\ref{fig:tree_analytical}). 

Our main contribution is introducing a novel representation and associated operators to enable neural geometry processing. We: 
\begin{enumerate}[(i)]
\item introduce spherical neural surfaces as a representation to encode genus-0 surfaces seamlessly as overfitted networks; 
\item compute surface normals, First and Second Fundamental Forms, curvature, continuous Laplace Beltrami operators without unnecessary discretization; 
\item approximate the smallest spectral modes of the Laplace Beltrami operator on spherical neural surfaces;
\item show illustrative applications towards scalar field manipulation and neural mean curvature flow; and
\item demonstrate the robustness and accuracy ours on different forms of input and quantitatively compare against alternatives. 
\end{enumerate}

\section{Related Works}

\paragraph*{Neural Representations.}

Implicit neural representations, such as signed distance functions~(SDFs) and occupancy networks, have recently been popularized to model 3D shapes. Notably, DeepSDF~\cite{deepSDF:19} leverages a neural network to represent the SDF of a shape, enabling continuous and differentiable shape representations that can be used for shape completion and reconstruction tasks; Davies et al.~\cite{DBLP:journals/corr/abs-2009-09808} use neural networks to overfit to individual SDFs;  
Occupancy Networks~\cite{occupancyNet:18} learn a continuous function to represent the occupancy status of points in space, providing a powerful tool for 3D shape modeling and reconstruction from sparse input data.
Explicit neural representations, on the other hand, use neural networks to directly predict structured 3D data such as meshes or point clouds. Mesh R-CNN~\cite{meshRCNN:19} extends Mask-RCNN to predict 3D meshes from images using a voxel-based representation followed by mesh refinement with a graph convolution network operating on the meshes' vertices and edges. In the context of point clouds,  PointNet~\cite{qi2016pointnet} and its variants are widely used as backbones for shape understanding tasks such as shape classification and segmentation.
Hybrid representations combine the strengths of both implicit and explicit methods. For example, Pixel2Mesh~\cite{pixel2mesh:18} generates 3D meshes from images by progressively deforming a template mesh using a graph convolutional network, integrating the detailed geometric structure typical of explicit methods with the continuous nature of implicit representations. More relevant to ours is AtlasNet~\cite{groueix2018} that models surfaces as a collection of parametric patches, balancing flexibility and precision in shape representation, and Neural Surface Maps~\cite{morreale2021neural}, which are overfitted to a flattened disc parametrization of surfaces to enable surface-to-surface mapping. Similarly to SNS, Explicit Neural Surfaces~\cite{walker2023} were proposed in the context of 3D reconstruction from images, but the authors did not yet explore the properties of the differentiable surface representation.

\paragraph*{Estimating Differential Quantities} Traditionally, researchers have investigated how to carry over differential geometry concepts~\cite{docarmo1976differential} to surface meshes, where differential geometry does not directly apply because mesh faces are flat, with all the `curvature' being at sharp face intersections. Taubin~\shortcite{taubin:95} introduced several signal processing operators on meshes. Meyer et al.~\shortcite{Meyer2002DiscreteDO} used averaging Voronoi cells and the mixed Finite-Element/Finite-Volume method; 
\cite{cazals2005estimating} used osculating jets; 
while \cite{Goes2020DiscreteDO} used discrete differential geometry~\cite{ddg:06:course} to compute gradients, curvatures, and Laplacians on meshes. Recently, researchers have used learning-based approaches to `learn' differential quantities like normals and curvatures on point clouds~\cite{GuerreroEtAl:PCPNet:EG:2018,ben2019nesti,pistilli2020point}.

More related to ours is the work by Novello et al.~\shortcite{novell_implicitNeuralSurface_22}, who represent surfaces via implicit functions and analytically compute differential quantities such as normals and curvatures. We refer to this method as `i3d' in our comparison~\ref{fig:tree_analytical}. Later, Novello et al. proposed a related method, to  evolve implicit functions~\cite{novello2023neural}, and \cite{mehta2022level} describes an alternative approach. With a similar motivation to Novello et al., Chetan et al.~\shortcite{chetan2023accurate}
fit    
local polynomial patches to obtain more accurate derivatives from pre-trained hybrid neural fields; they also use these as higher-order constraints in the neural implicit fitting stage. A similar idea was explored by Bednarik et al.~\shortcite{bednarik2020}, as a regularizer for the AtlasNet setup, by using implicit differential surface properties during training. The work by Yang and colleagues~\shortcite{yang2021geometry} is closest to ours in motivation - they also mainly focus on normal and curvature estimates but they additionally demonstrate how to perform feature smoothing and exaggeration using the local differential quantities in the target loss functions and thus driving the modification of the underlying neural fields.
In Section~\ref{sec:results}, we compare our estimates against those obtained by others~\cite{novell_implicitNeuralSurface_22,yang2021geometry}. 
However, we go beyond normal and curvature estimates and focus on operators such as surface gradient, surface divergence and Laplace-Beltrami operators for processing scalar and vector fields defined on the surfaces.

\paragraph*{Laplace Beltrami Operators on Meshes}
The Laplace-Beltrami operator~(LBO) is an indispensable tool in spectral geometry processing, as it is used to analyze and manipulate the intrinsic properties of shapes.  Spectral mesh processing~\cite{spectralMesh:10} leverages the eigenvalues and eigenfunctions of the LBO for tasks such as mesh smoothing and shape segmentation. Many subsequent efforts have demonstrated the use of LBO towards shape analysis and understanding (e.g., shapeDNA~\cite{shapeDNA:06}, global point signatures~\cite{rustamov2007GPS}, heat kernel signature~\cite{sun2009concise}), eventually culminating into the functional map framework~\cite{ovsjanikov2012functional}. An earlier signal processing framework~\cite{taubin:95} has been revisited with the LBO operator~\cite{spectralMesh:10} for multi-scale processing applications (e.g., smoothing and feature enhancement). 
Although the Laplace-Beltrami operator on triangular meshes or, more generally, on polyhedral meshes, can be computed, these operators depend on the underlying discretization of the surface (i.e., placement of vertices and their connectivity). See \cite{LBcourse:23} for a discussion. In the case of triangular meshes, the commonly used discretizations are the Uniform Laplace Operator and the cotan LBO. In Section~\ref{sec:results}, we compare our neural LBO with the cotan discretization and discuss their relative merits.

\section{Spherical Neural Surfaces}
\begin{figure}[h!]
    \centering
    \includegraphics[width=\columnwidth]{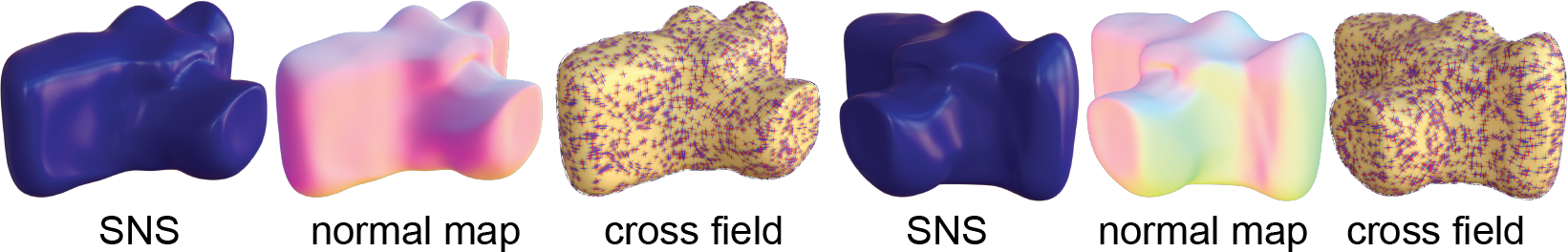}
    \mycaption{Handling Neural Implicits}{
    Spherical Neural Surfaces can be generated from other neural representations, such as DeepSDF. Here, we have produced an SNS from a DeepSDF representation for a camera, by overfitting the SNS to a low-resolution marching-cubes mesh reconstruction and then finetuning this SNS to better align with the SDF.
    }
\label{fig:SNS_deepSdf}
\vspace*{-.15in}
\end{figure}

We introduce a novel neural representation for continuous surfaces, called a `Spherical Neural Surface'~(SNS). An SNS is a Multi-Layer Perceptron~(MLP) $S_\theta:\,\mathbb{R}^3 \rightarrow \mathbb{R}^3$, trained such that its restriction to the unit sphere ($\mathbb{S}^2 \subset \mathbb{R}^3$) is a continuous parametrization of a given surface. In terms of notation, we use $S_\theta$ or $S$ as shorthand for the set $S_\theta (\mathbb{S}^2)$. In our current work, we primarily focus on creating SNSs from input triangle-meshes and analytically-defined surfaces, but there is potential to create SNSs from other representations such as pointclouds, SDFs and NeRFs. For example, Figure~\ref{fig:SNS_deepSdf} shows initial results for generating an SNS from a neural SDF.

\begin{figure}[b!]
    \centering
    \includegraphics[width=\columnwidth]{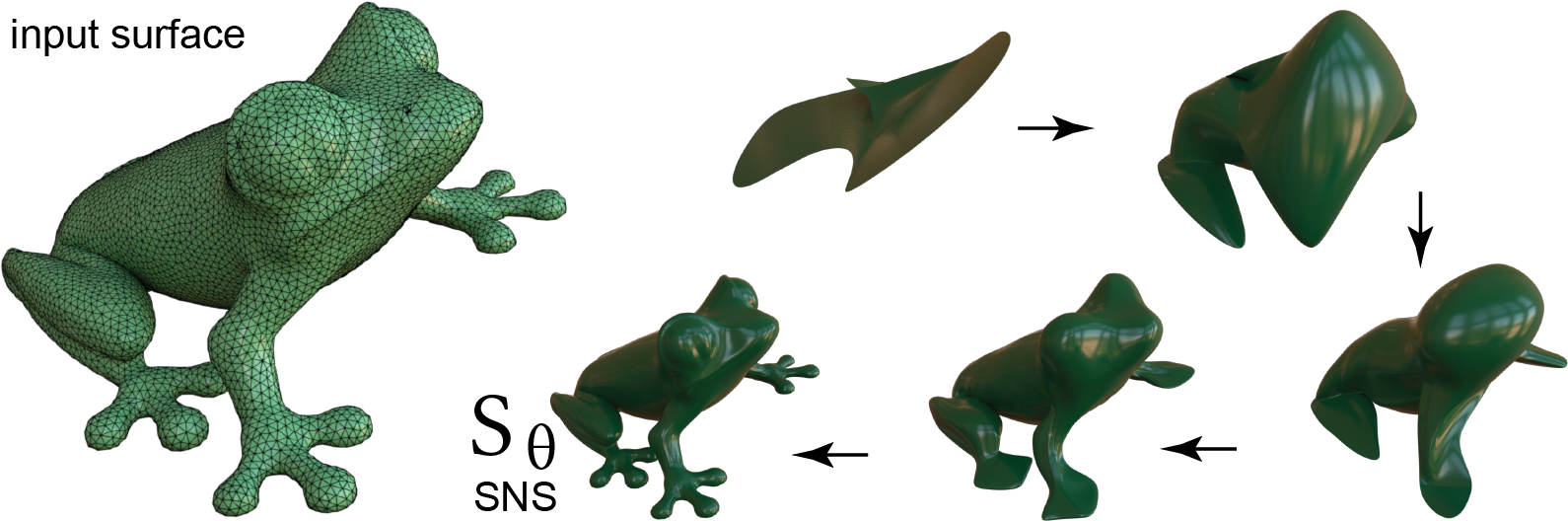}
    \mycaption{Spherical Neural Surface}{Given an input surface --- a mesh in this case --- we progressively overfit an MLP to a spherical mesh-embedding, to represent the surface as $S_\theta$. The loss term is simply the MSE between the ground truth and predicted surface positions, plus a (scaled) normals regularization term (Equation \ref{eq:loss_normal}).
    }
\label{fig:sns_over_iterations}
\end{figure}

\subsection{Overfitting an SNS to a Triangle Mesh}
Given a genus-0 manifold triangle mesh, we first find an injective embedding of the mesh vertices onto the unit sphere, using a spherical embedding algorithm~\cite{schmidt2023surface,praun:sphericalParam:02}. We extend the embedding to a continuous embedding by employing barycentric coordinates, so that every point $p_i$ on a (curved) triangle on the sphere corresponds to a unique point $q_i$ on a (flat) triangle on the target surface. We overfit the network $S_\theta$ to this parametrization by minimizing the MSE between the ground truth and predicted surface positions:
\begin{equation}
\mathcal{L}_{MSE} := \frac{1}{N}  \sum\limits_{i=1}^N \left \|  S_\theta(p_i) - q_i   \right \|^2  .
\end{equation}
Further, to improve the fitting, we encourage the normals derived from the spherical map to align with the normals of points on the mesh, via a regularization term, 
\begin{equation}
\mathcal{L}_{\textit{normal}} := \frac{1}{N}   \sum_ {i=1}^{N} \left \|  \mathbf{n}_{S_\theta}({p_i}) - \mathbf{n}_{\textit{mesh}}({q_i}) \right \|^2  = \frac{2}{N}   \sum_ {i=1}^{N} (1-\cos{\alpha_i}).
\label{eq:loss_normal}
\end{equation}
In this expression, $\mathbf{n}_{\textit{mesh}}({q_i})$ is the (outwards) unit-normal on the mesh at $q_i$, and $\mathbf{n}_{S_\theta}({p_i})$ is the corresponding (outwards) unit-normal of the smooth parametrization at $S_\theta(p_i)$, which is derived analytically from the Jacobian of $S_\theta$ at $p_i$ (see Section~\ref{subsec:diffEstimates}); the angle $\alpha_i$ is the angle between $\mathbf{n}_{\textit{mesh}}({q_i})$ and $\mathbf{n}_{S_\theta}({p_i})$.
Figure~\ref{fig:sns_over_iterations} shows an example of an SNS over training stages. 

\subsection{Computing Differential Quantities using SNS}
\label{subsec:diffEstimates}
One of the advantages of Spherical Neural Surfaces as a geometric representation is that one can compute important quantities from continuous differential geometry - without any need for approximate numerical formulations or discretization. We just need to use the automatic differentiation functionality built into modern machine learning setups, and algebraically track some changes of variables, before applying the formulas from classical differential geometry.

For example, to compute the outwards-pointing normal $\mathbf{n}_{S}$, we first use \texttt{autodiff} to compute the $3 \times 3$ Jacobian of $S$, \begin{equation*} \mathbf{J}({S}) :=  \begin{bmatrix}\frac{
\partial{S}}{\partial{x}}, &  \frac{
\partial{S}}{\partial{y}}, &\frac{
\partial{S}}{\partial{z}}   \end{bmatrix},
\end{equation*}
and then turn this into a $3 \times 2$ Jacobian in local coordinates, by composing it with a $3 \times 2$ orthonormal matrix ($R$):
\begin{equation}
\mathbf{J}^{\textit{local}}_{S} = \mathbf{J}({S})\,R .
\end{equation}
\removed{The matrix R defines orthogonal unit vectors on the tangent plane
of the sphere at each point. }\added{The matrix $R$ maps vectors in 2D local coordinates to tangent vectors on the sphere at a particular point, and the matrix $\mathbf{J}(S)$ maps tangent vectors on the sphere to tangent vectors on the surface $S$.} If $u$ and $v$ are the local coordinates on the tangent plane of the sphere, then the columns of $\mathbf{J}^{\textit{local}}_{S}$ are equal to the partial derivatives $S_u$ and $S_v$ respectively:
\begin{equation}
  \mathbf{J}^{\textit{local}}_{S} =  \begin{pmatrix}
\textbar & \textbar \\
S_u & S_v \\
\textbar & \textbar \\
\end{pmatrix},
\end{equation}
%
%
and these vectors lie in the tangent plane.
The normalized cross product of $S_u$ and $S_v$ is the outward-pointing unit normal:
\begin{equation}
\mathbf{n}_{S} = \frac{S_u \times S_v}{\left \| S_u \times S_v  \right \|}.
\end{equation}
In terms of standard spherical polar coordinates ($\theta$ and $\phi$), we chose to use the following rotation matrix:
\begin{equation}
    R = 
    \begin{pmatrix}
    \cos \theta \cos \phi & -\sin \phi \\
    \cos \theta \sin \phi & \cos \phi \\
    -\sin \theta & 0 \\
\end{pmatrix}.
\end{equation}
\added{The trigonometric quantities here are simple functions of $x$,$y$ and $z$.}


The matrix $R$ is an orthonormal version of the usual spherical polar coordinates Jacobian. We avoid degeneracy at the poles by selectively re-labeling the $x$, $y$, and $z$ coordinates so that $\theta$ is never close to zero or $\pi$. (This is equivalent to choosing a different chart for the points within some radius of the poles.)\par
Now, from $\mathbf{J}^{\textit{local}}_{S}$, we can simply write the First Fundamental Form of the sphere-to-surface map as:
\begin{equation}
\mathbf{I}_{S}=
  \begin{pmatrix}
E & F \\
F & G \\
\end{pmatrix} = (\mathbf{J}^{\textit{local}}_{S})^{T}\,\mathbf{J}^{\textit{local}}_{S}.
\end{equation}
To compute the Second Fundamental Form, we first compute the second partial derivatives of the parametrization with respect to the local coordinates, and then find the dot products of the second derivatives with the normal, $\mathbf{n} = \mathbf{n}_{S_\theta}$. The second derivatives are:
%
%
\begin{equation}
S_{uu} = \mathbf{J}(S_u)S_u, \quad 
S_{uv} = \mathbf{J}(S_u)S_v, \quad 
S_{vv} = \mathbf{J}(S_v)S_v. 
\end{equation}
We can now express the Second Fundamental Form as:
\begin{equation}\mathbf{II}_{S}=
 \begin{pmatrix}
e & f \\
f & g \\
\end{pmatrix} =  \begin{pmatrix}
S_{uu}\cdot\mathbf{n} & S_{uv}\cdot\mathbf{n} \\
S_{uv}\cdot\mathbf{n} & S_{vv}\cdot\mathbf{n} \\
\end{pmatrix}.
\end{equation}
In terms of the Fundamental Forms, the Gauss curvature is 
\begin{equation}
K = \frac{eg-f^2}{EG-F^2},
\end{equation}
and the mean curvature is 
\begin{equation}
H = \frac{Eg-2Ff+Ge}{2(EG-F^2)}.
\end{equation}

\begin{figure}[t!]
    \centering    \includegraphics[width=\columnwidth]{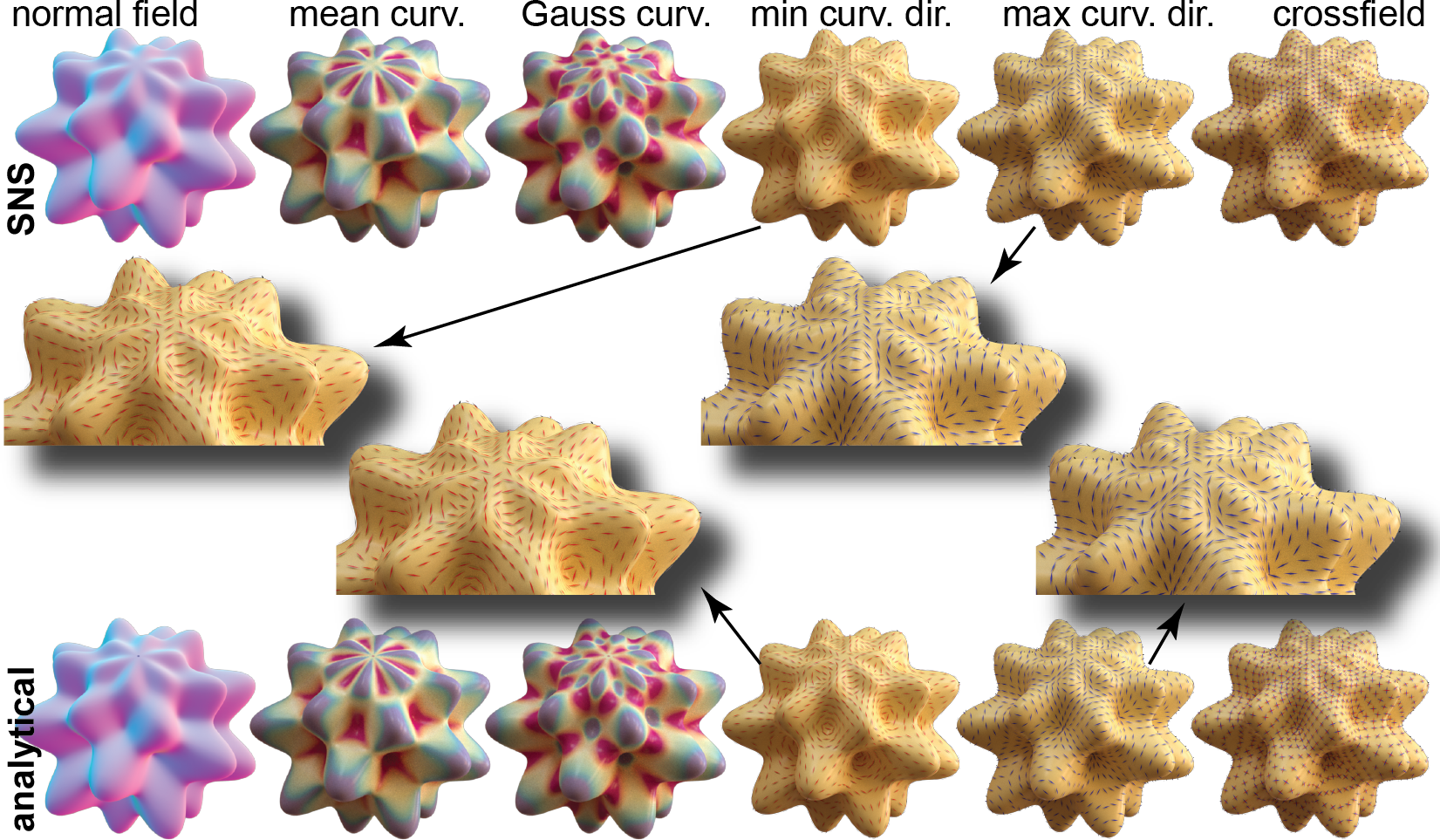}
    \mycaption{First and Second Fundamental Forms}{We show an analytically defined surface, \texttt{star} (bottom) and an SNS~(top row) that has been overfitted to a mesh of the surface (with 40962 vertices). We show various differential quantities computed symbolically using Matlab~(bottom) and computed on the SNS using our method~(top).
    Our results align very closely with the analytically-computed results.
    }

\label{fig:differentialEstimates_analytical}
\end{figure}

\begin{figure*}[t!]
    \centering
    \includegraphics[width=\textwidth]{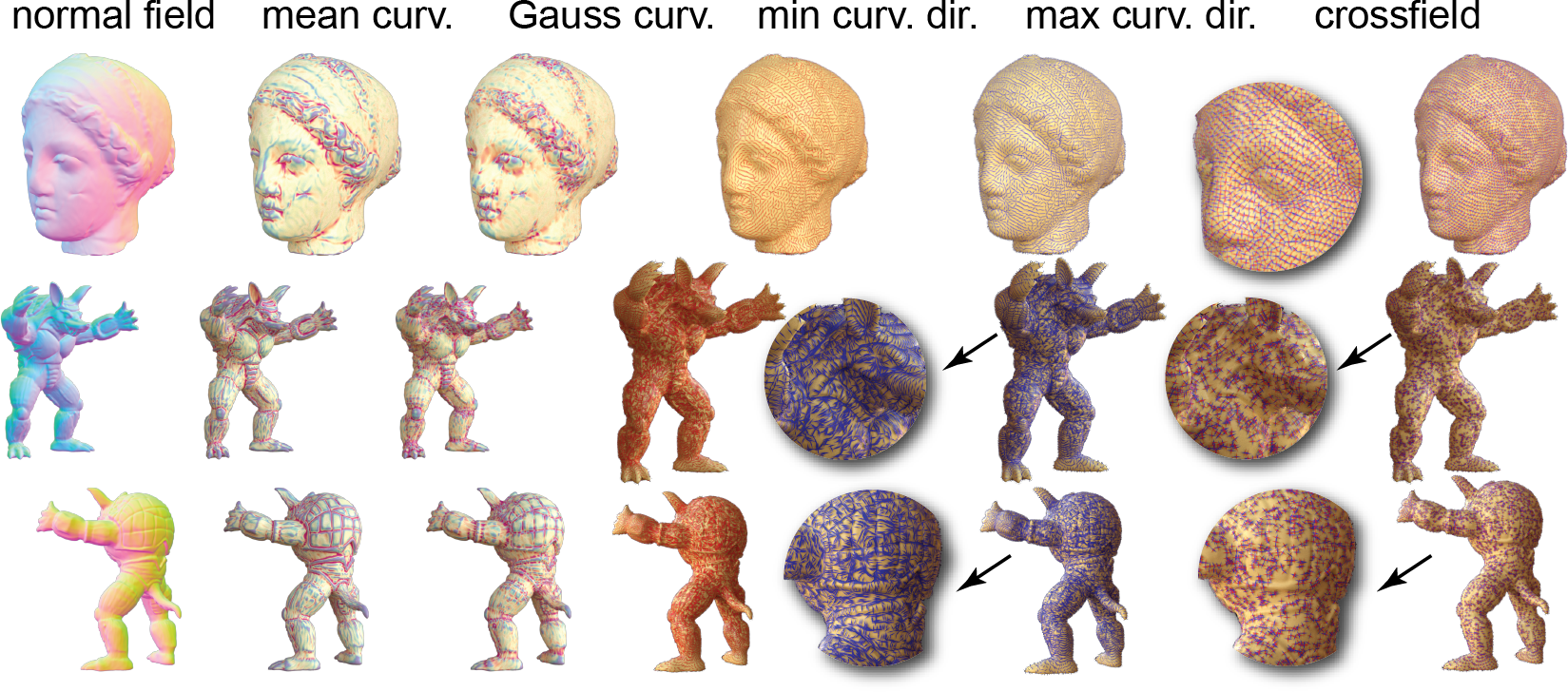}
    \mycaption{First and Second Fundamental Forms}{An SNS represents a smooth, seamless parametrization of a genus-0 surface, and we can compute the partial derivatives of this parametrization using automatic differentiation. We compute the normals and the First Fundamental Form of the parametrization from the first derivatives. Subsequently, computing the second derivatives enables us to construct the Second Fundamental Form of the parametrization. Hence, we compute Mean Curvature, Gaussian Curvature, and Principal Curvature Directions at any point without any unnecessary discretization error. \added{To render scalar fields, we took samples at the 2.6 million vertices of an icosphere.}}
\label{fig:differentialEstimates_real}
\end{figure*}

Furthermore, the solutions to the quadratic equation\\ $\text{det} (\mathbf{I}_S - \lambda \mathbf{II}_S)=0$ give us explicit expressions for the principal curvatures, that we then use to find the principal curvature directions directly on the Spherical Neural Surface. In Figure~\ref{fig:differentialEstimates_analytical}, we compare our differential estimates against analytical computations for an analytic surface; in Figure \ref{fig:differentialEstimates_real} we also show curvature estimates and principal curvature directions on other SNSs overfitted to non-analytic surfaces (ground truth values are not available). In Section~\ref{sec:results}, we evaluate the quality of our estimates and also evaluate how the estimates change as we fit SNSs to different mesh resolutions as proxies to the same underlying (continuous) surface.

\section{\removed{Mathematical Background}\added{Laplace-Beltrami Operator}}

This section briefly summarizes the relevant background for continuous surfaces that we need to develop a computation framework for Laplace-Beltrami and spectral analysis using SNS (Section~\ref{sec:SNS_spectral}).

\subsection{Laplace-Beltrami Operator on Smooth Surfaces}
The (Euclidean) Laplacian $\Delta$ of a scalar field $f:\mathbb{R}^n\rightarrow\mathbb{R}$ is defined as the \textit{divergence} of the \textit{gradient} of the scalar field. Equivalently, $\Delta f$ is the sum of the unmixed second partial derivatives of $f$: 
\begin{equation}
\Delta f := \nabla \cdot \nabla f =  \sum _{i=1}^{n}\frac{\partial^2 f }{\partial  x_i ^2}. 
\end{equation}

The Laplace-Beltrami operator, or \textit{surface Laplacian}, is the natural generalization of the Laplacian to scalar fields defined on curved domains. If $\Sigma \subset \mathbb{R}^3$ is a regular surface and $f:\Sigma \rightarrow \mathbb{R}$ 
is a scalar field, 
then the Laplace-Beltrami operator on $f$ is defined as the \textit{surface divergence} of the \textit{surface gradient} of $f$, and we denote this by $\Delta_\Sigma f$ to emphasize the dependence on the surface $\Sigma$. Thus, 
\begin{equation}
\Delta_\Sigma f := \nabla_\Sigma \cdot \nabla_\Sigma f.
\label{eqn:LBdefn}
\end{equation}
The surface gradient of the scalar field $f$ (written $\nabla_\Sigma f$), and the surface divergence of a vector field ${\mathbf{F}}:\Sigma \rightarrow \mathbb{R}^3$ (written $\nabla_\Sigma \cdot \mathbf{F}$), can be computed by smoothly extending the fields to fields $\tilde{f}$ and $\tilde{\mathbf{F}}$ respectively, which are defined on a neighborhood of $\Sigma$ in $\mathbb{R}^3$. 

The surface gradient is defined as the orthographic projection of the Euclidean gradient of $\tilde{f}$ into the local tangent plane:
\begin{equation}
\nabla_\Sigma f = \nabla \tilde{f} - (\nabla \tilde{f} \cdot \mathbf{n})\mathbf{n}, \label{eqn:surface_grad}
\end{equation}
where $\mathbf{n}$ is the unit surface normal at the point. \added{One can show that this definition is independent of the choice of extension.}
The surface divergence of the vector field $\mathbf{F} $
is defined as
\begin{equation}
\nabla_\Sigma \cdot \mathbf{F} =           \nabla \cdot \tilde{\mathbf{F}} - \mathbf{n}^T {\mathbf{J}}(\tilde{\mathbf{F}})  \mathbf{n}, 
\end{equation}
where $\mathbf{J}(\tilde{\mathbf{F}}) :=  \begin{bmatrix}\frac{
\partial{\tilde{\mathbf{F}}}}{\partial{x_1}}, &  \frac{
\partial{\tilde{\mathbf{F}}}}{\partial{x_2}}, &\frac{
\partial{\tilde{\mathbf{F}}}}{\partial{x_3}}\\
\end{bmatrix} $ is the Jacobian of $\mathbf{F}$.
%
%
Since $\mathbf{J}(\tilde{\mathbf{F}})  \mathbf{n}$ is the directional derivative of $\tilde{\mathbf{F}}$ in the normal direction, then $\mathbf{n}^T \mathbf{J}(\tilde{\mathbf{F}})  \mathbf{n} = \mathbf{J}(\tilde{\mathbf{F}})  \mathbf{n} \cdot \mathbf{n} $ can be thought of as the contribution to the three-dimensional divergence of $\tilde{\mathbf{F}}$ that comes from the normal component of $\tilde{\mathbf{F}}$, and by ignoring the contribution from the normal component, we get the two-dimensional `surface divergence'. Although these expressions depend a-priori on the particular choice of extension, the surface gradient and surface divergence are in fact well-defined (i.e., any choice of extension will give the same result).\par

By expanding out the definitions of surface gradient and surface divergence, we can derive an alternative formula for the surface Laplacian ($\Delta_{\Sigma} f$) in terms of the (Euclidean) Laplacian ($\Delta \tilde{f}$), the gradient ($\nabla \tilde{f}$) and the Hessian ($\mathbf{H}(\tilde{f}) = \mathbf{J}(\nabla \tilde{f})^{T}$) as
\begin{equation}
\Delta_\Sigma f = \Delta \tilde{f} - 2H \nabla \tilde{f} \cdot \mathbf{n} -\mathbf{n}^T \mathbf{H}(\tilde{f}) \mathbf{n}.
\label{eqn:LB_meancurv}
\end{equation}
The dependence on the surface is captured by the normal function $\mathbf{n}$, and the mean curvature $H$. (Refer to \cite{reilly1982soap,xu2003eulerian} for a derivation.)
In the case when $\tilde{f}$ is a `normal extension' --- i.e., it is locally constant in the normal directions close to the surface --- then the second and third terms disappear, so that $\Delta_\Sigma f$ is consistent with $\Delta \tilde{f}$.\par
If we substitute in the coordinate function $\mathbf{x}$ for $\tilde{f}$, then the first and third terms disappear, and we are left with the familiar equation
\begin{equation}
\Delta_\Sigma \mathbf{x} = - 2H \mathbf{n}.
\end{equation}
This formula is often used in the discrete setting to compute the mean curvature from the surface Laplacian.\par
\added{As in Section \ref{subsec:diffEstimates}, these formulae are easily implemented for functions defined on an SNS. All that is required to implement Equation \ref{eqn:LB_meancurv} is that we have access to an extension $\tilde{f}$ and its first and second derivatives. Therefore it can be applied when ${f}$ is (the restriction to the surface of) a closed-form function on the embedding space (e.g., Figures \ref{fig:LapBelt_robustness_meshing} and \ref{fig:LapBelt_robustness_meshing_igea}), or when $f$ and $\tilde{f}$ are defined by a network, allowing us to use \texttt{autodiff} (see Section \ref{sec:spectrum}).}


We have provided Equation~\ref{eqn:LB_meancurv} as an alternative formula to the standard definition (Equation~\ref{eqn:LBdefn}) because where a field evolves on a domain that remains constant, it allows us to precompute the mean curvature $H$ and normal $\mathbf{n}$, and then only the Euclidean quantities ($\nabla \tilde{f}$, $\mathbf{H}(\tilde{f})$, and $\Delta \tilde{f}$) need to be updated. In Figure~\ref{fig:scalarLB} (2\textsuperscript{nd} and 3\textsuperscript{rd} row) we demonstrate both methods and there is no visible difference.

\subsection{Spectrum of the Laplace-Beltrami Operator} \label{sec:spectrum}
Given two scalar functions, $f$ and $g$, defined on the same regular surface $\Sigma$, their $L^2$ inner product is
\begin{equation}
\left < f, g \right >_{L^2 (\Sigma)} := \int_\Sigma fg \,dA \,.
\end{equation}
Then, the $L^2$ norm of a scalar function can be expressed as,
\begin{equation}
\left \| f \right \|_{L^2(\Sigma)} = \sqrt{\left < f, f \right >_{L^2 (\Sigma)} } = \sqrt { \int_\Sigma |f|^2 \,dA } \,.
\end{equation}
The Laplace Beltrami operator, $\Delta_\Sigma$, is a self-adjoint linear operator with respect to this inner product. This means that there is a set of eigenfunctions of $\Delta_\Sigma$ that form an orthonormal basis for the space $L^2 (\Sigma)$ - the space of `well-behaved' scalar functions on $\Sigma$ that have a finite $L^2$-norm. This basis is analogous to the Fourier basis for the space of periodic functions on an interval.

In the discrete mesh case, the eigenfunctions of the Laplace-Beltrami operator are computed by diagonalizing a sparse matrix \cite{spectralMesh:10,LBcourse:23} based on the input mesh. In the continuous case, however, we no longer have a finite-dimensional matrix representation for $\Delta_\Sigma$. Instead, we exploit a functional analysis result that describes the first $k$ eigenfunctions of $\Delta_\Sigma$ as the solution to a minimization problem, as described next.

First, we define the Rayleigh Quotient of a scalar function $f$ to be the Dirichlet energy of the scalar function, divided by the squared $L^2$-norm of the function:
\begin{equation}
Q_\Sigma(f) :=  \frac{\left \| \nabla_\Sigma f \right \|_{L^2 (\Sigma)}^2}{\left \| f \right \|^2 _{L^2(\Sigma)}}.
\label{eq:rayleigh}
\end{equation}
Then, the first $k$ eigenfunctions of $\Delta_\Sigma$ are the minimizers of the Rayleigh Quotient, as given by, 
\begin{equation*}
\Psi_0, \Psi_1, ..., \Psi_{k-1} = \text{argmin}_{\Psi_0^*, ..., \Psi_{k-1}^*} \sum_{i=0}^{k-1} Q_\Sigma(\Psi_i^*)
\end{equation*}
\begin{equation}
\textit{such that}\quad \left < \Psi_i , \Psi_j \right >_{L^2(\Sigma)}=0 \quad \text{for all} \quad i\neq j .
\end{equation}
We use $\Psi_0$ to denote the first eigenfunction, which is always a constant function. The constraint states that the eigenfunctions are orthogonal inside the inner-product space $L^2 (\Sigma)$.\par

In addition, the Rayleigh Quotient of the eigenfunction $\Psi_i$ is the positive of the corresponding (negative) eigenvalue, i.e., 
\begin{equation}
    \Delta_\Sigma \Psi_i(x) = -Q_\Sigma(\Psi_i) \Psi_i(x) \quad \forall \, x \in \Sigma.
\end{equation}
Physically, the eigenfunctions of $\Delta_\Sigma$ represent the fundamental modes of vibration of the surface $\Sigma$. The Rayleigh Quotient of $\Psi_0$ is zero. As the Rayleigh quotient increases, the energy increases, and the eigenfunctions produce higher frequency patterns.

\begin{figure}[b!]
    \centering
    \includegraphics[width=\columnwidth]{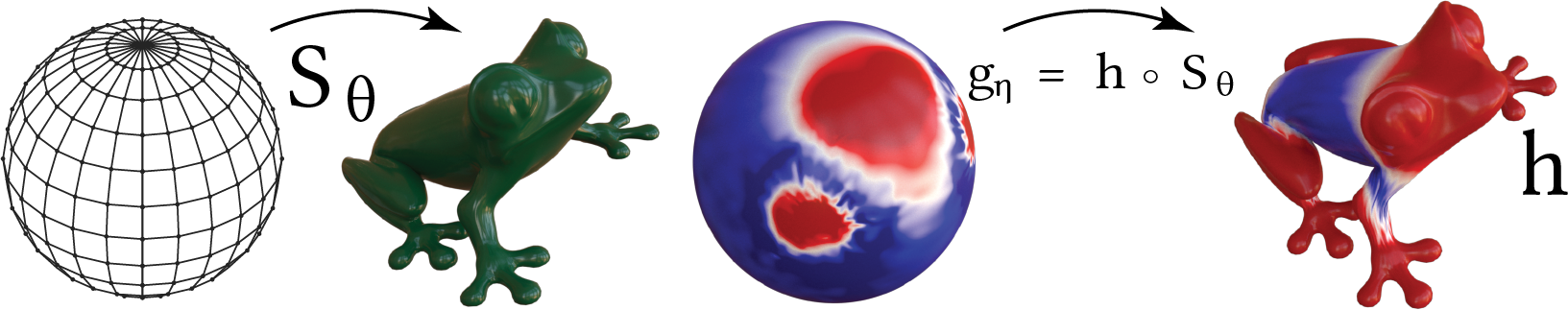}
    \mycaption{Encoding Scalar Fields on Neural Surfaces}{
We encode scalar fields on the sphere (and corresponding extensions into $\mathbb{R}^3$) as MLPs $g_\eta:\mathbb{R}^3 \rightarrow \mathbb{R}$, with parameters $\eta$. Because $S_\theta$, considered as a parametrization, is bijective, then we can implicitly define any smooth scalar field $h$ on the surface in terms of one of the functions $g_\eta$ (restricted to $\mathbb{S}^2$). We visualize the one-to-one correspondence between scalar fields defined on the surface, and scalar fields defined on the sphere, by using vertex-colours to display the fields.
}
    \label{fig:scalarIllustration}
\end{figure}

\begin{figure}[t!]
    \centering
    \includegraphics[width=\columnwidth]{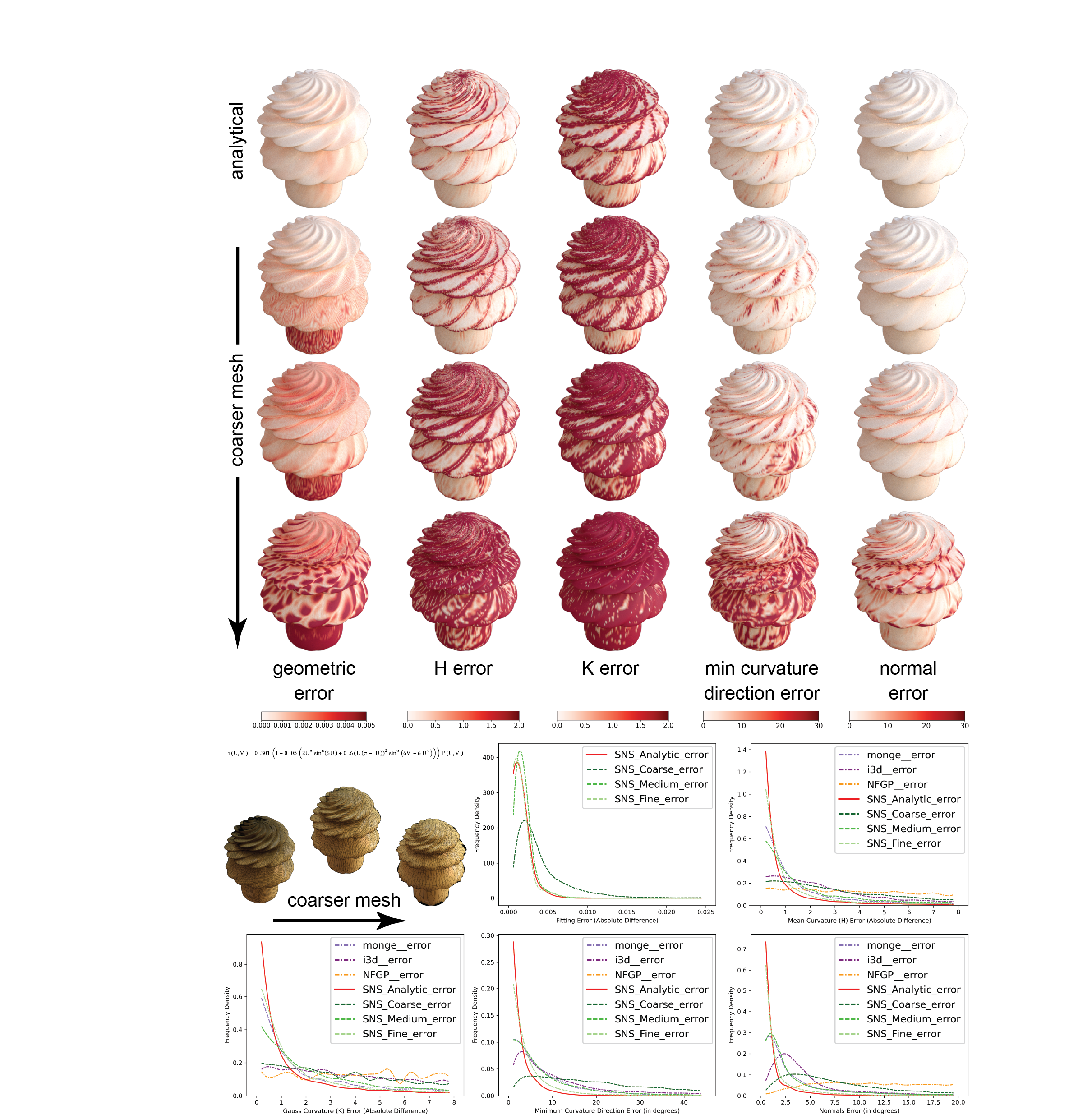}
    \mycaption{Accuracy of Differential Quantity Estimation}{As shown by the colourmaps, the quality of the estimation of differential quantities (mean curvature, Gauss curvature, minimum curvature direction and normals) decreases gradually as mesh-resolution decreases. We show results for an SNS overfitted to analytical parametrization, with 102400 sample points (top), and SNSs fitted to meshes with 40962, 10242 and 2562 vertices. (The errors are sampled at a much higher resolution, at 655362 points.) The level of error for the finest level mesh is close to the level of error for the `analytical' SNS and we argue that this demonstrates the `interpolation' capability of our method: it shows that fitting an SNS to an appropriately high-resolution mesh successfully interpolates the underlying surface (even for meshes where the analytic parametrization is not available). We also show plots of the error on the various differential quantities, compared against three additional methods (when functions are available): Monge-fitting, NFGP and i3d. A minimum curvature direction function was not available for NFGP.
    }
\label{fig:tree_analytical}
\end{figure}

\section{Spectral Analysis using SNS}
\label{sec:SNS_spectral}
To find the \textit{continuous} eigenmodes of the Laplace-Beltrami operator on an SNS, we require a continuous and differentiable representation for scalar functions defined on the surface. In our framework, we represent
smooth scalar fields on 
\added{the sphere}
$\mathbb{S}^2$ by MLPs, $g_\eta$, from $\mathbb{R}^3$ to $\mathbb{R}$, which are parametrized by $\eta$. We only ever evaluate the MLP $g_\eta$ on $\mathbb{S}^2 \in \mathbb{R}^3$, but because the \added{network's} domain is $\mathbb{R}^3$ then it defines an extension of the scalar field, and this allows us to compute \added{the Euclidean gradient} $\nabla g_\eta$. Then, if $S_\theta:\mathbb{R}^3 \rightarrow \mathbb{R}^3$ is an SNS, any smooth scalar field $h$ \added{whose domain is the surface defined by $S_\Theta$} can be defined implicitly by the equation
\begin{equation}
h  \circ S_\theta = g_\eta.
\label{eqn:implicit}
\end{equation}
Applying the chain rule, we can compute the gradient of $h$:
\begin{equation}
\nabla h = (\mathbf{J}(S_\theta)^T)^{-1} \nabla g_{\eta},
\end{equation}
which allows us to compute $\nabla_{S_\theta} h$ (Equation~\ref{eqn:surface_grad}) and $Q_{S_\theta}(h)$ (Equation~\ref{eq:rayleigh}) \textit{without} explicitly computing $h$.

To optimize for a (smooth) scalar function $g_{\eta_k}$ so that $h$ approximates the $k$th non-constant eigenfunction of the Laplace-Beltrami operator, we use gradient descent to optimize the weights, $\eta_k$. Specifically, we use a combination of two loss terms:
\begin{equation}
\mathcal{L}_{Rayleigh} := Q_S(h)  \quad \text{and} \quad 
\mathcal{L}_{ortho} := \sum_{i=0}^{k-1} \left <  h, h_i  \right >_{L^2(S)}^2 ,
\end{equation} where we sequentially compute the eigenfunctions, and $h_i$ is the $i$\textsuperscript{th} smallest eigenfunction. Recall that we represent each such eigenfunction using a dedicated MLP (except for $h_0$, which is the constant eigenfunction). The loss term $\mathcal{L}_{ortho}$ encourages the scalar field $h$ to be orthogonal to all previously-found eigenfunctions (including the constant eigenfunction). The quantity $Q_{S_\theta} (h)$ depends on the Jacobian of $S_\theta$, through $\nabla_{S_\theta} h$ and $\nabla h$. However, since the surface does not change during optimization and $\theta$ stays fixed, we precompute $(\mathbf{J}(S_\theta)^T)^{-1}$. 
Note that, if we were to optimize over the subspace of the unit-norm functions in $L^2(\Sigma)$, then we could replace the Rayleigh Quotient with the Dirichlet energy. \removed{However, it was a more natural design choice for us to parametrize the entire space $L^2(\Sigma)$ and optimize the Rayleigh Quotient, rather than strictly constrain the functions to have unit-norm and minimize the Dirichlet Energy.}\added{However, it is difficult to construct a function, based on an MLP, that produces only unit-norm functions (in the $L^2$ sense). In initial experiments, we used a  unit-norm loss in combination with the Dirichlet Energy and orthogonality loss, but we observed some robustness issues, such as a tendency for the functions to collapse to zero. We found that robustness increased when we replaced the Dirichlet Energy by the Rayleigh Quotient, and we think that this is the Rayleigh Quotient, by cancelling out the effect of magnitude, is less in opposition to the unit-norm loss than the Dirichlet energy is. The Rayleigh Quotient design choice greatly reduces the importance of the unit-norm loss, but we retained it as a regularizer, to prevent functions from collapsing to zero.  }

Although functions $h$ are not required to have unit norm, we used a regularization term to prevent the $L^2$-norm of $h$ from either exploding or vanishing:
\begin{equation}
   \mathcal{L}_{reg} := (\left \| h \right \|_{L^2(S)} - 1)^2.
\end{equation}
The combined loss term is expressed as, 
\begin{equation}
   \mathcal{L} = \mathcal{L}_{Rayleigh} + \lambda_{ortho}\mathcal{L}_{ortho} + \lambda_{reg}\mathcal{L}_{reg}. 
\end{equation}
We choose large initial values for the coefficients because we find that this improves stability, and then we reduce the coefficients so that the Rayleigh Quotient is the dominant term during the later stages of optimization (see Section \ref{sec:results} for details).

Finally, since we cannot exactly compute the integrals for the inner products in these loss terms, we approximate the inner products using Monte-Carlo sampling. Specifically, we approximate, 
\begin{equation}
\left < h, h_i \right >_{L^2 (S)} \approx \frac{Area(S)}{N}\sum_{j=1}^N h(q_j)h_i(q_j),
\end{equation}
where $\{q_j\} _{1 \leq j \leq N}$ is a set of uniformly distributed points on the surface $S_\theta$, corresponding to points  $\{p_j\} _{1 \leq j \leq N}$ on the sphere; $h(q_j)$ and $ h_i(q_j)$ are evaluated by applying scalar field MLPs $g_\eta$ and $g_{\eta_i}$ to $p_j$. (In the overfitting stage, we normalize all our Spherical Neural Surfaces to have an area equal to $4\pi$.) For the Monte Carlo step, we generate $N$ uniformly distributed points on the surface $S_\theta$ via the following steps (rejection sampling):
\begin{enumerate}[(i)]
    \item Generate $M \gg N_{target}$ samples ($p_j$) from the 3D normal distribution $\mathcal{N}(\mathbf{\mu}=\mathbf{0},{\Sigma}= \mathbf{I})$. Normalize, to unit norm, to produce a dense uniform distribution of points on the sphere $\mathbb{S}^2 \subset \mathbb{R}^3$.
    \item Compute the local area distortion ($d_j = \sqrt{E_jG_j-F_j^2}$), for each point $p_j$, using the First Fundamental Form.
    \item Perform rejection sampling, such that the probability of keeping point $p_k$ is equal to $ \frac{d_k}{\sum_{j=1}^{M} d_j}N_{target}$.
    \item Push the samples $p_j$ (on the sphere) through the map $S_\theta$.
\end{enumerate}
Following this process, the \textit{expected} number of sample points will equal $N_{target}$ --- a specified parameter --- however, the actual number of sample points ($N$) may vary.

\begin{figure}[t!]
    \centering
    \includegraphics[width=\columnwidth]{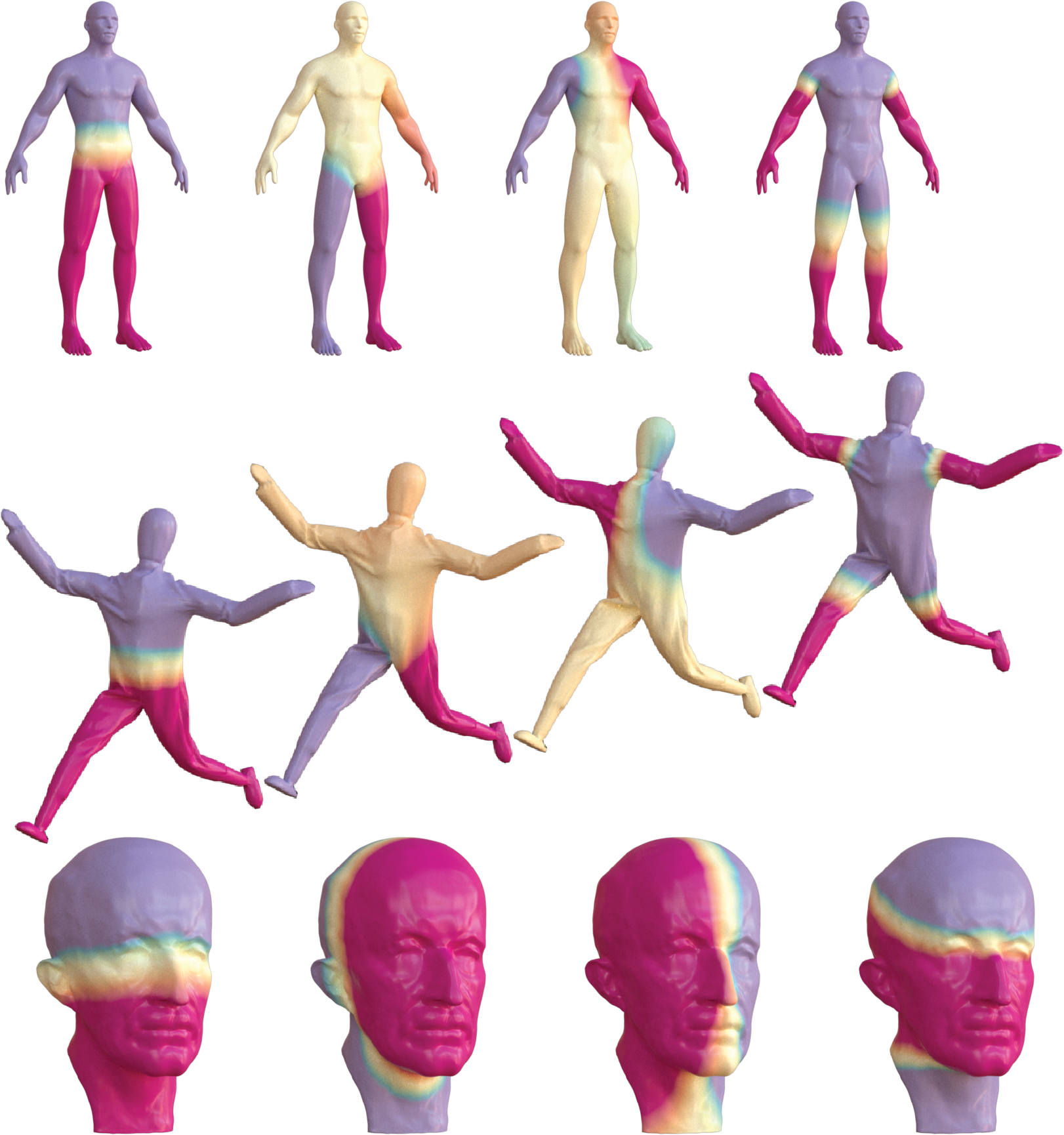}
    \mycaption{Spectral Modes on Neural Surfaces}{
    We propose how to compute the lowest few spectral modes of the neural Laplace Beltrami operator $\Delta_{S_\theta}$ via minimization of the Rayleigh quotient, without requiring any unnecessary discretization. Here we show the lowest few spectral modes of different SNSs. The lowest spectral mode is constant, and not included in this figure. Each of the spectral modes is represented by a dedicated MLP (see Equation \ref{eqn:implicit}). 
    }
\label{fig:spectral_modes}
\end{figure}

\begin{figure}[t!]
    \centering
\includegraphics[width=\columnwidth]{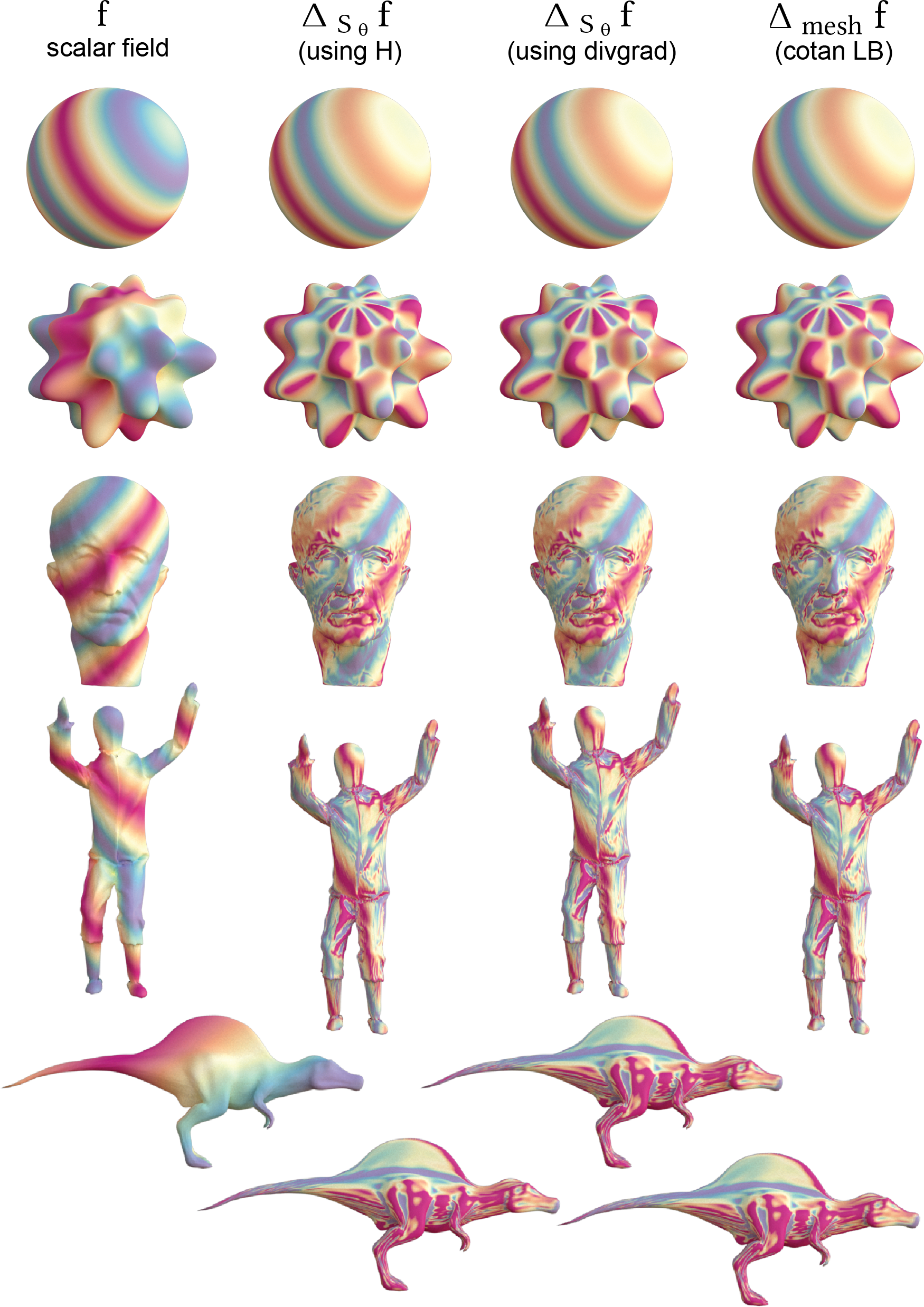}
    \mycaption{Laplace Beltrami on Neural Scalar Fields}{We present two ways to compute Laplace Beltrami operators on any smooth scalar field, $f$, defined on a neural surface $S_\theta$~(top): first, using mean curvature estimates (top-middle)~(Equation~\ref{eqn:LB_meancurv}) and second, using the divergence of gradient definition~(divgrad) (bottom-middle)~(Equation~\ref{eqn:LBdefn}). We represent scalar fields on surfaces using dedicated MLPs. (Bottom)~For comparison, we also show results using the cotan LBO (with a lumped mass matrix) on a dense mesh, whose vertices are the image of the vertices in a dense icosphere mesh under the transformation $S_\theta$. 
  }
\label{fig:scalarLB}
\end{figure}

\begin{figure*}[t!]
    \centering
    \includegraphics[width=\textwidth]{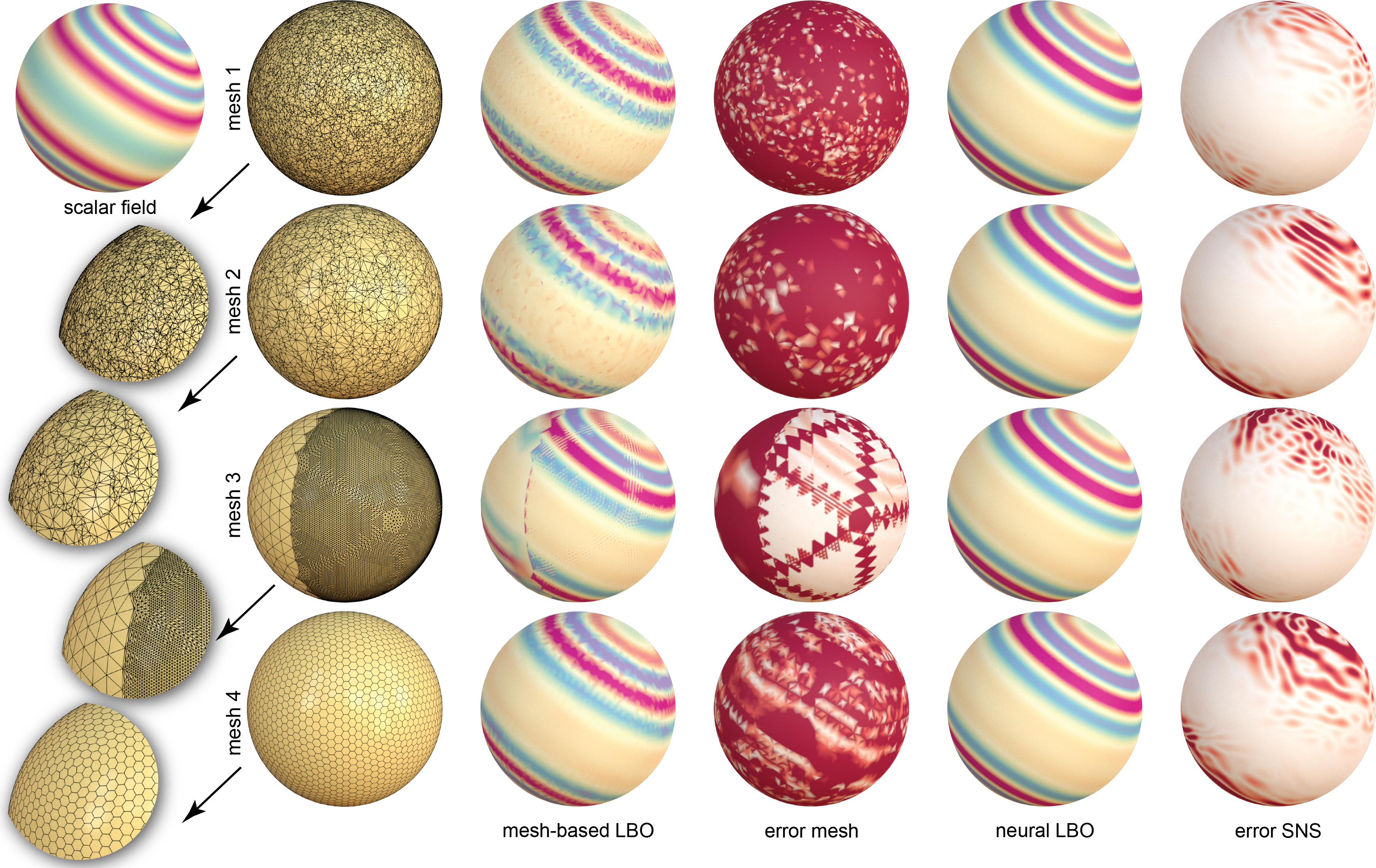} 
    \mycaption{Robustness of Neural LBO (Analytic)}{We demonstrate that our neural LBO is largely unaffected by differences in mesh density and mesh quality on the starting mesh. While the common mesh-based LBOs are sensitive to the quality and the density of the mesh, the neural LBO is more consistent when applied to a scalar field defined on an SNS that is generated from different meshes of the same surface. The mesh-based LBO is the cotan LBO for the triangle-meshes and the virtual-refinement Laplace~\cite{bungePolyLaplace2020} for polygonal meshes. The scalar field is analytic (a variable frequency sine-wave) on a sphere (for which the mean-curvature is one and the normals are trivial), because this allows us to compute hessian, gradient, normals and mean curvature analytically, so that we can compare against the ground truth LBO using equation \ref{eqn:LB_meancurv}. Please refer to the supplemental for further details. The error colourmap runs from white (no error) to red (higher error).
     }
\label{fig:LapBelt_robustness_meshing}
\end{figure*}

\begin{figure}[t!]
    \centering
    \includegraphics[width=\columnwidth]{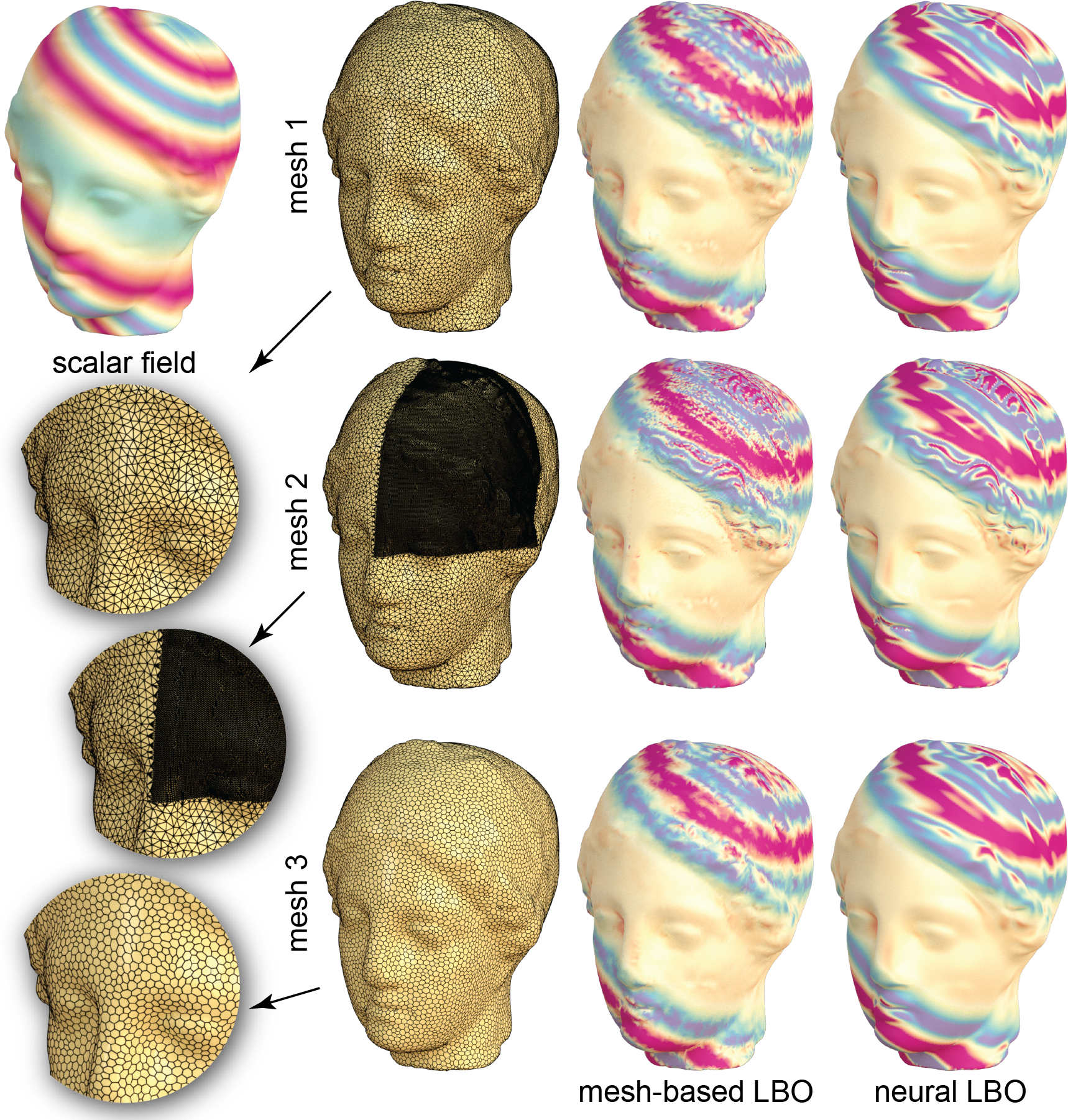} 
    \mycaption{Robustness of Neural LBO (Non-Analytic)}{For non-analytic surfaces, we do not have access to a ground-truth LBO, because the precise underlying surface geometry is ambiguous. However, we show that the neural LBO is more consistent than mesh-based LBOs when fitted to different meshes of the same surface. This is because the smooth geometry of the SNS remains quite consistent when fitted to different meshes of the same surface, and the neural LBO is defined everywhere on the smooth surface - not just at the mesh-vertices. The neural LBO on mesh 2 does display some visible differences, however this is because the hair of Igea contains some high-frequency details. This is a signal which was lost in the lower-resolution meshes, so it is not an artefact or noise. }
\label{fig:LapBelt_robustness_meshing_igea}
\end{figure}

\section{Evaluation}
\label{sec:results}

\paragraph*{Dataset}
We evaluate our method on various meshes of the unit sphere, various meshes of the Igea model, and four analytic shapes. In order to evaluate the robustness of our method, we included several `bad' meshes (e.g. with many thin triangles) and unusual meshes such as those with a very sharp contrast in triangle size across different regions,  and two polygonal meshes. (See Figures~\ref{fig:LapBelt_robustness_meshing} and \ref{fig:LapBelt_robustness_meshing_igea}.)\par
The four analytic shapes, which we designed for this experiment, are: \texttt{mushroom}, \texttt{flower}, \texttt{bobble}, and \texttt{star}. In Figure \ref{fig:tree_analytical}, we show results on the \texttt{mushroom} only. Refer to the supplemental, for the analytic parametrizations of all four shapes, and for additional results on the \texttt{flower}, \texttt{bobble}, and \texttt{star}. The advantage of using analytic shapes for evaluation is that we can compute the ground truth for any derived geometric quantity (such as curvature). We used symbolic Matlab to compute the ground truth differential quantities, given the analytic parametrization. We provide text files, in the supplemental, which contain the symbolic expressions.

\paragraph*{Evaluation Metrics}
To evaluate the quality of the surface representation, we measure:
\begin{enumerate}[(i)]
\item The distance from each point on the reconstructed surface, to the analytic surface (geometric error).
\item The angle deviation of the surface normal at each (reconstructed) surface point, compared to the ground truth surface normal at the the closest analytic surface point.
\item The angle deviation of the minimum curvature direction at each (reconstructed) surface point, compared to the ground truth minimum curvature direction at the the closest analytic surface point.
\item The absolute error of the Gaussian curvature and Mean curvature at each (reconstructed) surface point, compared to the ground truth curvatures at the the closest analytic surface point.
\end{enumerate}

For the Laplace Beltrami operator, we compare our LBO to other LBOs when applied to a frequency-varying sine-wave type scalar field, and in the case of a sphere, we compute the absolute error with respect to the analytic LBO (see supplemental for details).

\paragraph*{Comparison Methods}

We compare the SNS against three alternative methods for estimating geometric quantities. (These methods do not support LBO computation or spectral analysis.)
\begin{itemize}
\item \textit{Monge-fitting~\cite{cgal:pc-eldp-24a}}] is a classical mesh-based method that uses osculating jets to approximate differential quantities of the underlying curved surface. We used the official implementation available via CGAL. 
\item \textit{NFGP~\cite{yang2021geometry}} is a method with similar motivation to ours, in the sense that they fit a neural representation to a mesh and compute differential quantities in order to perform edits in the shape. However, they use a deep-implicit representation, so their differential quantities additionally depend on the representation's adherance to the Eikonal property, whereas we use an explicit representation.
\item \textit{i3D~\cite{novell_implicitNeuralSurface_22}} is also a deep implicit method that computes differential quantities, and it improves the overfitting process itself by encouraging the curvature of the deep implicit to match the computed curvature of the input mesh. Because our method, NFGP, and Monge fitting do not require the curvature information from the mesh, we did not provide the mesh-curvature information to this method either and we found that the method suffered as a result.
\end{itemize}

\paragraph*{(i) What is the effect on the estimates when we fit the SNS to different resolution meshes of an analytic shape?}

In Figure~\ref{fig:tree_analytical}, we fit SNSs to analytic shapes (\texttt{mushroom}), and plot the absolute error of derived geometric quantities (against the ground truth, which is computed from the analytic parametrization using symbolic Matlab). We fit SNSs to a coarse, medium and fine mesh (with 2562, 10242 and 40962 vertices, respectively) and, for reference, we also fit an SNS directly to the analytic parametrization (without meshing).

As expected, the accuracy of the derived quantities is highest when we fit the SNS directly to the analytic parametrization, and the accuracy of the mesh-fitted SNSs approaches the analytically-fitted SNS as the mesh resolution increases. Since the accuracy of the SNS fitted to the fine-resolution mesh (40962 vertices) is consistently close to the accuracy of the analytically-fitted SNS, we argue that fitting the SNS to a reasonably fine mesh is a good way to find a smooth approximation of an underlying smooth surface, when no analytic parametrization is available. We also show the distribution of errors for one of the shapes (\texttt{mushroom}), for SNSs fitted to each resolution of the mesh, and for the three comparison methods: Monge-fitting, NFGP and i3D.\par

\paragraph*{(ii) How do the position, normals and curvature estimates of SNS compare to those from other methods?}
In Figure~\ref{fig:tree_analytical}, alongside the SNS errors, we show the absolute errors of computing the same geometric quantities on the fine-resolution mesh, using three alternative methods. The first method is Monge-fitting, which fits a piecewise polynomial surface to the mesh and allows us to compute the geometric quantities on the vertices. The other two methods (i3d and NFGP) are deep implicit methods, which fit a neural signed-distance function (a deepSDF) to the mesh. The i3D method has the option to use mesh-curvature information to improve the fitting, but for consistency, we did not provide this, because the other neural methods (ours, and NFGP) do not require additional mesh-curvature information. We use the CGAL implementation for Monge-fitting, and the original author -implementations for the two deep-implicit methods. The geometric quantities on a deepSDF are computed using automatic differentiation. (Although this appears at first glance to be similar to computing geometric quantities on an SNS, the computations are different due to deepSDF being an implicit representation while SNS is explicit. In contrast to SNS, deepSDFs use a soft constraint enforced by an Eikonal loss term, without which the deepSDF can only be used for mesh-reconstruction and not for the estimation of geometric quantities.)

We observe that given only a fine-resolution mesh, the most accurate method is consistently the SNS, except on the \texttt{flower} shape where it is Monge-fitting. The second most accurate method appears, on average, to be the Monge-fitting, but impressively the medium-resolution SNS is sometimes able to perform similarly well, showing that SNS has can robustly interpolate from limited geometric data.

On all shapes, i3D performed worse than the fine-resolution SNS and the Monge-fitting.
The NFGP method produces consistently poor results. We believe that this method struggles to capture finer surface details (e.g. on the \texttt{mushroom}) and thin structures (e.g. on the \text{flower}), at least with the default settings. However, we were limited by compute resources and it is possible that better results could be achieved given a longer training time (please see the `NFGP Note' in the Supplemental).
Observations of the reconstructed meshes show that neither of the deep implicit methods capture the input geometry well, leading to poor results in the estimation of geometric quantities.\par

\paragraph*{(iii) How consistent is the SNS LBO, when we fit an SNS to different meshes of the same shape?}

In Figure \ref{fig:LapBelt_robustness_meshing} and 
\ref{fig:LapBelt_robustness_meshing_igea},    we put a test-function (a variable-frequency sine-wave scalar field) on two different surfaces: the sphere and Igea, respectively. We fit the SNS to various different meshings of these shapes, and in each case compute the result of the applying the SNS LBO to the scalar field. To fit an SNS to a polygonal mesh, we first convert to a triangle mesh by splitting the faces.\par
For reference, we also show the result of applying a mesh-based LBO to the scalar field (sampled at the vertices). For triangle meshes, we use the cotan LBO (with vertex area as $1/3$ the one-ring area, and a lumped mass matrix). For polygonal meshes, we use the Virtual Refinement Laplace operator from Polygon Laplacian Made Simple~\cite{bungePolyLaplace2020}, with the default Area Minimizer setting.\par

Visually, the SNS LBO produces more consistent and less noisy results across different meshings. This is in part because the SNS LBO is discretization-free, so we can sample at any point (not just vertices of the original mesh), and partly because the SNS LBO is able to leverage information from the derivatives of the scalar field (which is always available when the scalar field is analytic, or neural) whereas the mesh-based methods \textit{cannot} use this additional available information.

\paragraph*{(iv) How does the SNS LBO compare to the ground truth LBO, when the analytic operator is available?}
In the case of a sphere, we know the mean curvature and normals analytically at every point, so we can use Equation \ref{eqn:LB_meancurv} to compute the ground-truth result of applying the LBO to the (analytically-defined) scalar field. We show this result in Figure \ref{fig:LapBelt_robustness_meshing} , and also display the error maps of the mesh-based LBOs and the neural LBO. Clearly, the error of the neural LBO is smaller than the error of the mesh-based LBO, and the error is smallest where the scalar field is low-frequency.

\paragraph*{(v) Neural LBO on scalar fields on non-analytic surfaces.}
To evaluate the accuracy of our Laplace-Beltrami operator on SNSs, we generated random scalar fields using sine waves, and visually compared the results of applying three different forms of the LBO, on five different surfaces (see Figure~\ref{fig:scalarLB}). To compute $\Delta_{S_\theta} f$ using the mean curvature formulation (Equation \ref{eqn:LB_meancurv}), we computed the gradient and Hessians of the scalar field analytically and combined these with the SNS estimates of the normals and mean curvature. We also computed $\Delta_{S_\theta} f$ using the `divergence of gradient' definition (Equation \ref{eqn:LBdefn}) - in this version, all of the 
differential calculations were carried out by \texttt{autograd}. We show the results using vertex-colours, on a dense icosphere mesh.\par
Finally, we computed the result of applying the cotan LBO to the scalar field (sampled at the vertices of the dense icosphere mesh). At this resolution, all three versions of the LBO are extremely close, however, we notice that the cotan estimate on the \texttt{star} (row 2) contains a small discontinuity, not present in the SNS estimates. From a smooth shape and a smooth scalar field, the resulting $\Delta_{S}$ should be  smooth, which is the case for the SNS estimates.

\begin{figure}[h!]
    \centering
    \includegraphics[width=\columnwidth]{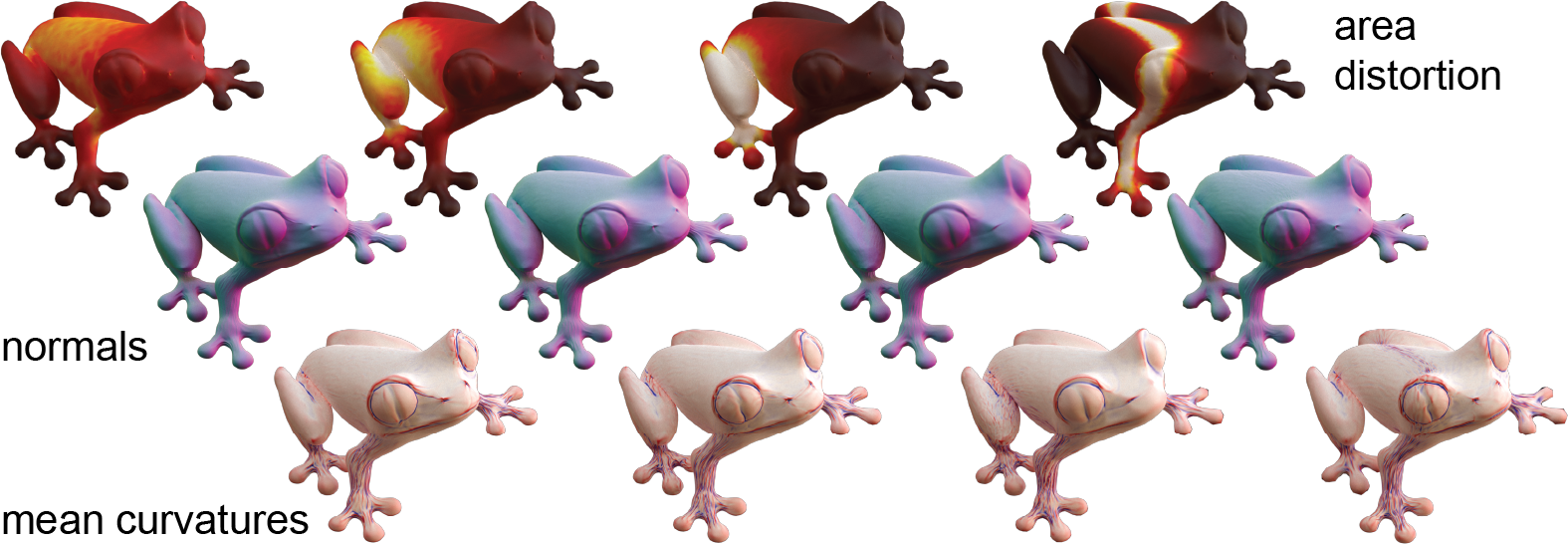}
    \caption{\added{We show area-distortion colourmaps for four different spherical embeddings of the frog mesh, and demonstrate the effect of different patterns of distortion on the accuracy of fitting, and normals and curvature estimates. 
    }}
    \label{fig:distortion}
\end{figure}

\added{
\paragraph*{(vi) What is the effect of different levels of distortion in the spherical mesh embedding?}
{Figure \ref{fig:distortion} shows area-distortion colourmaps for four different spherical embeddings of the frog mesh. Spherical embeddings were computed by performing explicit deformations of the mesh-embedding produced by multi-resolution embedding~\cite{schmidt2023surface}. The distortion colourmap is shown on the original mesh. The area-distortion at a single vertex is computed as the one-ring area on the sphere mesh divided by the corresponding one-ring area on the surface mesh.
The second and third rows show the SNS trained on each parametrization, with colouring based on the normals and the mean curvature. We see that first order estimates (normals) are more consistent than second order estimates (curvature). We also see that, although the geometry is close to the input surface in all cases, artefacts such as `ripples' occur more frequently in white areas, where the MLP maps a large area on the sphere to a small area on the surface.} }

\paragraph*{Implementation Details}
All of our networks in our experiments were trained on Nvidia GeForce P8 GPUs.
The SNS network is a Residual MLP with input and output of size three, and eight hidden layers with 256 nodes each. The networks used to represent scalar fields (in the eigenfunction and heat flow experiments) are very small Residual MLPs with input size three, output size three and two hidden layers with 10 nodes each. (To make the output value a scalar, we take the mean of the three output values.) \added{This rather small network-size appears to be sufficient to represent the smallest eigenfunctions, however it might be necessary to add more nodes if we aim to represent highly complex scalar fields.} For both scalar fields and SNS, we use \texttt{softplus} as the activation function. We use the RMSProp optimizer with a learning rate of $10^{-4}$ and a momentum of 0.9.
We trained each SNS for up to 20,000 epochs (eight to ten hours), and fixed the normal regularization coefficient to be $10^{-4}$. The time per epoch increases slightly with the number of vertices in the mesh, but for simpler shapes (such as the sphere) the optimization converges in fewer epochs.

\paragraph*{Can our optimization process reproduce eigenfunctions of the LBO when the analytic solution is known?}
In the eigenfunction experiment, we used the initial parameter values $\lambda_{ortho}=10^3$ and $\lambda_{unit}=10^4$ and then we reduced the coefficients linearly to $\lambda_{ortho}=1$ and $\lambda_{unit}=10^2$, over 10,000 epochs. These settings prevent the scalar field from collapsing to the zero-function during the initial stages of optimization, whilst allowing the Rayleigh Quotient to dominate the loss in the later stages. We optimized each eigenfunction for up to 40,000 epochs (approximately six hours on our setup).\par
We generate $M = 100,000$ points for the initial uniform distribution on the sphere, and we use $N_{target}=10,000$ in the rejection sampling process. The points stay fixed during each optimization stage. We believe that this is a possible reason why the Rayleigh Quotient computed on the training points is often slightly lower than the Rayleigh Quotient that is computed with a different, larger set of uniform samples, and a better sampling strategy might improve the accuracy of our optimization.

\begin{figure}[b!]
    \centering
\includegraphics[width=\columnwidth]{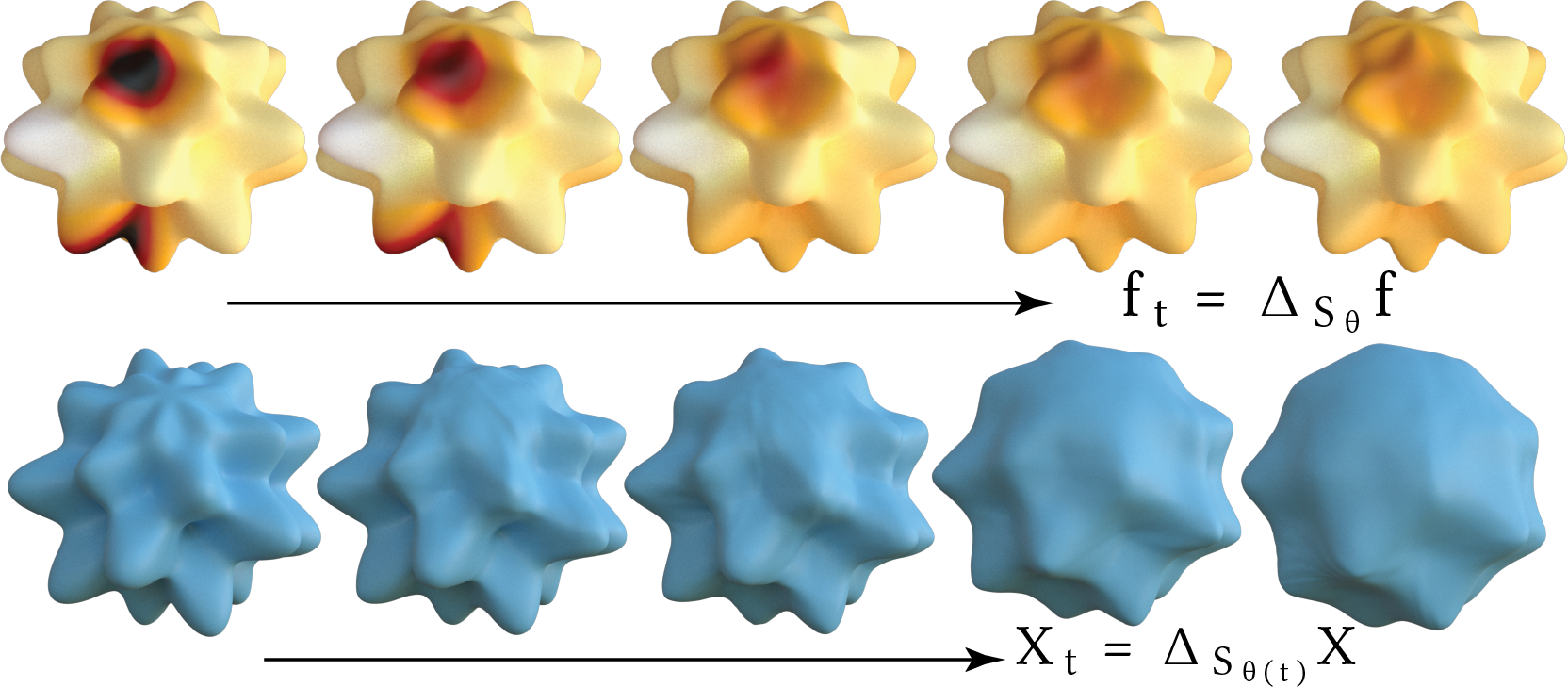}
    \mycaption{Heat Flow~(top) and Mean Curvature Flow~(bottom)}{
    (Top)~We can evolve a given scalar field, $f$, specified over the surface of an SNS, $S_\theta$, using the Partial Differential Equation~(PDE) $f_t = \Delta_{S_\theta} f$ (the heat equation). We represent the evolving scalar field as a small MLP, whose weights are `finetuned' for up to 100 epochs after every time step. Darker colours denote low values (cold) with lighter colours being high (hot). 
    (Bottom)~Taking advantage of the differentiable nature of our representation, we can also compute a mesh-free Mean Curvature Flow~(MCF) in which the Mean Curvature and normals are
updated at every iteration (the coordinate function evolves according to the PDE $X_t = \Delta_{S_{\theta(t)}} X$). We update the Spherical Neural
Surface using up to 100 finetuning steps, after every iteration of the flow. This formulation of MCF prevents
singularities from forming, without any special handling to prevent them~\cite{kazhdan2012meancurvature}. 
}
\label{fig:heat+meancurvature}
\end{figure}

\paragraph*{Heat Flow and Mean Curvature Flow}
 We can approximate the evolution of any initial scalar field $f$ on an SNS $S_\theta$, according to the process defined by the heat equation,
 $
  f_t = \Delta_{S_\theta} f.
$  In our implementation, we represent the field $f$ implicitly, using Equation \ref{eqn:implicit}. In each time step, we compute $\Delta_{S_\theta} f$ for the current scalar field $f$, using the divergence of gradient formula in Equation \ref{eqn:LBdefn}. Then we select a small value ($d=10^{-3}$) and we compute $f + d \Delta_{S_\theta}f$ at some fixed sample points (in Figure \ref{fig:heat+meancurvature}-left, we used 10,242 points). We then update the scalar field, by finetuning the network for up to 100 epochs to fit the new sample points, using a simple MSE loss. Qualitatively, the experiment aligns with our expectations. The scalar field gradually becomes more smooth over time, and cold areas warm up to a similar heat level as the surroundings.\par 
Similarly, we approximate a Mean Curvature Flow of an SNS by updating the SNS to $S_\theta -d H\mathbf{n}$ at each time step, and finetuning for up to 100 epochs, so that the surface evolves according to the PDE $X_t = \Delta_{S_{\theta (t)}} X$ The results in Figure \ref{fig:heat+meancurvature}-right show the spiky analytic shape becoming closer to a sphere, over 150 iterations.

\paragraph*{Neural Implicits}
To demonstrate robustness to the input representation, we provide an illustratory example of an SNS generated from a DeepSDF~\cite{deepSDF:19}, and show the normals and principal curvature directions on its SNS. We took a pretrained DeepSDF network and the optimized latent code for the camera shape \cite{comi2023deepsdf}. Then, as initialization, we overfitted an SNS to a coarse marching-cubes mesh. Then we took a set of samples on the SNS and projected them to deepSDF representation defined by the signed distance function (SDF), by moving them in the direction of the gradient of the SDF (which approximates the surface normal) by the signed distance at that point, and use them to finetune the SNS.

\section{Conclusion}

We have presented spherical neural surfaces as a novel representation for genus-0 surfaces. Our smooth and differentiable neural surface representation inherently supports the computation of continuous differential geometric quantities and operators such as normals, First and Second Fundamental Forms), surface gradients and 
surface divergence. We also presented how to compute a continuous Laplace Beltrami operator and its lowest spectral modes.

We demonstrate that SNSs, by avoiding the pitfalls of discretization, produce robust and consistent differential geometric estimates for surfaces both across meshing parameters such as sampling and vertex connectivity, and when compared to alternate neural and mesh-based surface representations.

\subsection*{Limitations and Future Work}

\paragraph*{Beyond Genus-0 Surfaces} Since our SNSs rely on a sphere for the parametrization, it limits us to genus-0 surfaces. We chose the sphere as a domain, because it allows us to process many common shapes without needing to `cut' them into a disc-topology (as shown in, e.g., Neural Surface Maps~\cite{morreale2021neural}). If we could choose from a range of canonical surfaces (e.g., torus), this would allow us to seamlessly process higher genus shapes and, possibly, to also produce neural surfaces without such high distortions. Another option would be to rely on local parametrizations, but then we would have to consider blending across overlapping parametrizations.

\paragraph*{Solving Variational Problems}
We are optimistic that the techniques demonstrated in our work have the potential to allow us to solve other variational problems, beyond the example of computing a spectral basis via energy minimization. We would like to know whether similar energy minimizations may be used on neural surface-paths, to compute geodesics, and on scalar- and vector-fields, to compute heat-flow/mean-curvature flow without fine-tuning the network at every step.

\paragraph*{Speed} A limitation of our current realization is its long running time. While we expect that better and optimized implementations would increase efficiency significantly, we also need to make changes to our formulation. Specifically, our current spectral estimation is sequential, which is not only slow, but leads to an accumulation of errors for later spectral modes. Hence, in the future, we would like to jointly optimize for multiple spectral modes. 

\paragraph*{SNS from Neural Representations} We demonstrated our setup mainly on mesh input and also on neural input in the form of deepSDF. In the future, we would like to extend our SNS to other neural representations in the form of occupancy fields or radiance fields. However, this will require locating and sampling points on the surfaces, which are implicitly encoded -- we need the representation to provide a projection operator. We would also like to support neural surfaces that come with level-of-detail. Finally, an interesting direction would be to explore end-to-end formulation for dynamic surfaces encoding temporal surfaces with SNS, and enabling optimization with loss terms involving first/second fundamental forms as well as Laplace-Beltrami operators (e.g., neural deformation energy).


\paragraph*{Acknowledgements}The authors wish to thank Johannes Wallner, for his constructive comments on an early version of this work, as well as Uday Kusupati and Karran Pandey, for proofreading. We thank \href{https://www.morenap.ca/}{Morena Protti}, for the amazing \href{https://www.thingiverse.com/morenap/designs}{treefrog mesh}. RW was supported by the Engineering and Physical Sciences Research Council (grant
number EP/S021566/1). NM was supported by Marie Skłodowska-Curie grant agreement No.~956585, gifts from Adobe, and the UCL AI Centre.
%

\bibliographystyle{eg-alpha-doi}  
\bibliography{SphericalNeuralSurfaces}

\newcommand{\etalchar}[1]{$^{#1}$}
\begin{thebibliography}{\uppercase{WMKG07}}

\bibitem[BAB23]{LBcourse:23}
\textsc{Bunge A., Alexa M., Botsch M.}:
\newblock Discrete {L}aplacians for general polygonal and polyhedral meshes.
\newblock In \emph{SIGGRAPH Asia 2023 Courses} (2023), Association for Computing Machinery.
\newblock URL: \url{https://dl.acm.org/doi/10.1145/3610538.3614620}.

\bibitem[BHKB20]{bungePolyLaplace2020}
\textsc{Bunge A., Herholz P., Kazhdan M., Botsch M.}:
\newblock Polygon {L}aplacian made simple.
\newblock In \emph{Computer Graphics Forum} (2020), vol.~39, Wiley Online Library, pp.~303--313.
\newblock URL: \url{https://onlinelibrary.wiley.com/doi/full/10.1111/cgforum.13931}.

\bibitem[BPG{\etalchar{*}}20]{bednarik2020}
\textsc{Bednarik J., Parashar S., Gundogdu E., Salzmann M., Fua P.}:
\newblock Shape reconstruction by learning differentiable surface representations.
\newblock In \emph{Proceedings IEEE Conf. on Computer Vision and Pattern Recognition (CVPR)} (2020).
\newblock \href {https://doi.org/10.1109/CVPR42600.2020.00477} {\path{doi:10.1109/CVPR42600.2020.00477}}.

\bibitem[BSLF19]{ben2019nesti}
\textsc{Ben-Shabat Y., Lindenbaum M., Fischer A.}:
\newblock Nesti-net: Normal estimation for unstructured 3{D} point clouds using convolutional neural networks.
\newblock In \emph{Proceedings IEEE Conf. on Computer Vision and Pattern Recognition (CVPR)} (2019), IEEE, pp.~10112--10120.
\newblock URL: \url{https://doi.org/10.48550/arXiv.1812.00709}.

\bibitem[Com23]{comi2023deepsdf}
\textsc{Comi M.}:
\newblock {DeepSDF}-minimal.
\newblock \url{https://github.com/maurock/DeepSDF/}, 2023.

\bibitem[CP05]{cazals2005estimating}
\textsc{Cazals F., Pouget M.}:
\newblock Estimating differential quantities using polynomial fitting of osculating jets.
\newblock \emph{Computer Aided Geometric Design 22}, 2 (2005), 121--146.
\newblock URL: \url{https://inria.hal.science/inria-00097582v1}.

\bibitem[CYW{\etalchar{*}}23]{chetan2023accurate}
\textsc{Chetan A., Yang G., Wang Z., Marschner S., Hariharan B.}:
\newblock Accurate differential operators for hybrid neural fields, 2023.
\newblock \href {http://arxiv.org/abs/2312.05984} {\path{arXiv:2312.05984}}.

\bibitem[dC76]{docarmo1976differential}
\textsc{do~Carmo M.~P.}:
\newblock \emph{Differential Geometry of Curves and Surfaces}.
\newblock Prentice-Hall, 1976.

\bibitem[dGBD20]{Goes2020DiscreteDO}
\textsc{de~Goes F., Butts A., Desbrun M.}:
\newblock Discrete differential operators on polygonal meshes.
\newblock \emph{ACM Transactions on Graphics TOG 39} (2020), 110:1 -- 110:14.
\newblock URL: \url{https://api.semanticscholar.org/CorpusID:221105828}.

\bibitem[DNJ20]{DBLP:journals/corr/abs-2009-09808}
\textsc{Davies T., Nowrouzezahrai D., Jacobson A.}:
\newblock Overfit neural networks as a compact shape representation.
\newblock \emph{CoRR abs/2009.09808} (2020).
\newblock URL: \url{https://arxiv.org/abs/2009.09808}, \href {http://arxiv.org/abs/2009.09808} {\path{arXiv:2009.09808}}.

\bibitem[DPSS06]{ddg:06:course}
\textsc{Desbrun M., Polthier K., Schröder P., Stern A.}:
\newblock Discrete differential geometry - an applied introduction.
\newblock In \emph{ACM {SIGGRAPH} course notes} (2006).

\bibitem[GFK{\etalchar{*}}18]{groueix2018}
\textsc{Groueix T., Fisher M., Kim V.~G., Russell B., Aubry M.}:
\newblock {AtlasNet: A Papier-M\^ach\'e Approach to Learning 3{D} Surface Generation}.
\newblock In \emph{Proceedings IEEE Conf. on Computer Vision and Pattern Recognition (CVPR)} (2018).
\newblock URL: \url{https://doi.org/10.48550/arXiv.1802.05384}.

\bibitem[GKOM18]{GuerreroEtAl:PCPNet:EG:2018}
\textsc{Guerrero P., Kleiman Y., Ovsjanikov M., Mitra N.~J.}:
\newblock {PCPNet}: Learning local shape properties from raw point clouds.
\newblock \emph{Computer Graphics Forum 37}, 2 (2018), 75--85.
\newblock \href {https://doi.org/10.1111/cgforum.13343} {\path{doi:10.1111/cgforum.13343}}.

\bibitem[GMJ19]{meshRCNN:19}
\textsc{Gkioxari G., Malik J., Johnson J.}:
\newblock Mesh {R-CNN}.
\newblock \emph{CoRR abs/1906.02739} (2019).
\newblock URL: \url{http://arxiv.org/abs/1906.02739}, \href {http://arxiv.org/abs/1906.02739} {\path{arXiv:1906.02739}}.

\bibitem[KSBC12]{kazhdan2012meancurvature}
\textsc{Kazhdan M., Solomon J., Ben-Chen M.}:
\newblock Can mean-curvature flow be made non-singular?, 2012.
\newblock \href {http://arxiv.org/abs/1203.6819} {\path{arXiv:1203.6819}}.

\bibitem[LZ10]{spectralMesh:10}
\textsc{L\'{e}vy B., Zhang H.~R.}:
\newblock Spectral mesh processing.
\newblock In \emph{ACM SIGGRAPH 2010 Courses} (2010), SIGGRAPH '10.

\bibitem[MAKM21]{morreale2021neural}
\textsc{Morreale L., Aigerman N., Kim V.~G., Mitra N.~J.}:
\newblock Neural surface maps.
\newblock In \emph{Proceedings of the IEEE/CVF Conference on Computer Vision and Pattern Recognition} (2021), pp.~4639--4648.
\newblock URL: \url{https://doi.org/10.48550/arXiv.2103.16942}.

\bibitem[MCR22]{mehta2022level}
\textsc{Mehta I., Chandraker M., Ramamoorthi R.}:
\newblock A level set theory for neural implicit evolution under explicit flows.
\newblock \emph{ECCV} (2022).

\bibitem[MDSB02]{Meyer2002DiscreteDO}
\textsc{Meyer M., Desbrun M., Schr{\"o}der P., Barr A.~H.}:
\newblock Discrete differential-geometry operators for triangulated 2-manifolds.
\newblock In \emph{International Workshop on Visualization and Mathematics} (2002).
\newblock URL: \url{https://api.semanticscholar.org/CorpusID:11850545}.

\bibitem[MON{\etalchar{*}}18]{occupancyNet:18}
\textsc{Mescheder L.~M., Oechsle M., Niemeyer M., Nowozin S., Geiger A.}:
\newblock Occupancy networks: Learning 3{D} reconstruction in function space.
\newblock \emph{CoRR abs/1812.03828} (2018).
\newblock URL: \url{https://doi.org/10.48550/arXiv.1812.03828}.

\bibitem[MST{\etalchar{*}}21]{nerf:21}
\textsc{Mildenhall B., Srinivasan P.~P., Tancik M., Barron J.~T., Ramamoorthi R., Ng R.}:
\newblock Ne{RF}: representing scenes as neural radiance fields for view synthesis.
\newblock \emph{Commun. ACM 65}, 1 (Dec 2021), 99–106.
\newblock URL: \url{https://doi.org/10.48550/arXiv.2003.08934}.

\bibitem[NDS{\etalchar{*}}23]{Nsampi2023NeuralFC}
\textsc{Nsampi N.~E., Djeacoumar A., Seidel H.-P., Ritschel T., Leimk{\"u}hler T.}:
\newblock Neural field convolutions by repeated differentiation.
\newblock \emph{ACM Trans. Graph. 42}, 6 (Dec 2023).
\newblock URL: \url{https://doi.org/10.1145/3618340}, \href {https://doi.org/10.1145/3618340} {\path{doi:10.1145/3618340}}.

\bibitem[NdSS{\etalchar{*}}23]{novello2023neural}
\textsc{Novello T., da~Silva V., Schardong G., Schirmer L., Lopes H., Velho L.}:
\newblock Neural implicit surface evolution.
\newblock \emph{ICCV} (2023).

\bibitem[NSS{\etalchar{*}}22]{novell_implicitNeuralSurface_22}
\textsc{Novello T., Schardong G., Schirmer L., {da Silva} V., Lopes H., Velho L.}:
\newblock Exploring differential geometry in neural implicits.
\newblock \emph{Computers and Graphics 108} (2022), 49--60.
\newblock URL: \url{https://doi.org/10.48550/arXiv.2201.09263}.

\bibitem[OBCS{\etalchar{*}}12]{ovsjanikov2012functional}
\textsc{Ovsjanikov M., Ben-Chen M., Solomon J., Butscher A., Guibas L.}:
\newblock Functional maps: a flexible representation of maps between shapes.
\newblock \emph{ACM Transactions on Graphics TOG 31}, 4 (2012), 1--11.
\newblock URL: \url{https://dl.acm.org/doi/10.1145/2185520.2185526}.

\bibitem[PC24]{cgal:pc-eldp-24a}
\textsc{Pouget M., Cazals F.}:
\newblock Estimation of local differential properties of point-sampled surfaces.
\newblock In \emph{{CGAL} User and Reference Manual}, 5.6.1~ed. {CGAL Editorial Board}, 2024.
\newblock URL: \url{https://doc.cgal.org/5.6.1/Manual/packages.html##PkgJetFitting3}.

\bibitem[PFS{\etalchar{*}}19]{deepSDF:19}
\textsc{Park J.~J., Florence P.~R., Straub J., Newcombe R.~A., Lovegrove S.}:
\newblock Deep{SDF}: Learning continuous signed distance functions for shape representation.
\newblock \emph{CoRR abs/1901.05103} (2019).
\newblock URL: \url{https://arxiv.org/abs/1901.05103}.

\bibitem[PFVM20]{pistilli2020point}
\textsc{Pistilli F., Fracastoro G., Valsesia D., Magli E.}:
\newblock Point cloud normal estimation with graph-convolutional neural networks.
\newblock In \emph{2020 IEEE International Conference on Multimedia \& Expo Workshops (ICMEW)} (2020), IEEE, pp.~1--6.
\newblock \href {https://doi.org/10.1109/ICMEW46912.2020.9105972} {\path{doi:10.1109/ICMEW46912.2020.9105972}}.

\bibitem[PH03]{praun:sphericalParam:02}
\textsc{Praun E., Hoppe H.}:
\newblock Spherical parametrization and remeshing.
\newblock \emph{ACM Transactions on Graphics TOG 22}, 3 (jul 2003), 340–349.
\newblock \href {https://doi.org/10.1145/882262.882274} {\path{doi:10.1145/882262.882274}}.

\bibitem[QSMG]{qi2016pointnet}
\textsc{Qi C.~R., Su H., Mo K., Guibas L.~J.}:
\newblock Point{N}et: Deep learning on point sets for 3{D} classification and segmentation.
\newblock \emph{arXiv preprint}.
\newblock URL: \url{https://doi.org/10.48550/arXiv.1612.00593}.

\bibitem[Rei82]{reilly1982soap}
\textsc{Reilly R.~C.}:
\newblock Mean curvature, the {L}aplacian, and soap bubbles.
\newblock \emph{The American Mathematical Monthly 89} (1982), 180--98.
\newblock URL: \url{https://api.semanticscholar.org/CorpusID:203043073}.

\bibitem[Rus07]{rustamov2007GPS}
\textsc{Rustamov R.}:
\newblock {L}aplace-{B}eltrami eigenfunctions for deformation invariant shape representation.
\newblock In \emph{Proceedings of the Symposium on Geometry Processing} (2007), Eurographics Association, pp.~225--233.
\newblock URL: \url{https://dl.acm.org/doi/10.5555/1281991.1282022}.

\bibitem[RWP06]{shapeDNA:06}
\textsc{Reuter M., Wolter F.-E., Peinecke N.}:
\newblock {L}aplace–{B}eltrami spectra as ‘shape-{DNA}’ of surfaces and solids.
\newblock \emph{Computer-Aided Design 38}, 4 (2006), 342--366.
\newblock Symposium on Solid and Physical Modeling 2005.
\newblock URL: \url{https://doi.org/10.1016/j.cad.2005.10.011}.

\bibitem[SC20]{monteCarlo:20}
\textsc{Sawhney R., Crane K.}:
\newblock Monte carlo geometry processing: a grid-free approach to {PDE}-based methods on volumetric domains.
\newblock \emph{ACM Trans. Graph. 39}, 4 (aug 2020).
\newblock URL: \url{https://dl.acm.org/doi/abs/10.1145/3386569.3392374}.

\bibitem[SOG09]{sun2009concise}
\textsc{Sun J., Ovsjanikov M., Guibas L.}:
\newblock A concise and provably informative multi-scale signature based on heat diffusion.
\newblock \emph{Computer Graphics Forum 28}, 5 (2009), 1383--1392.
\newblock URL: \url{https://onlinelibrary.wiley.com/doi/full/10.1111/j.1467-8659.2009.01515.x}.

\bibitem[SPK23]{schmidt2023surface}
\textsc{Schmidt P., Pieper D., Kobbelt L.}:
\newblock Surface maps via adaptive triangulations.
\newblock \emph{Computer Graphics Forum 42}, 2 (2023).
\newblock URL: \url{https://doi.org/10.1111/cgforum.14747}.

\bibitem[Tau95]{taubin:95}
\textsc{Taubin G.}:
\newblock A signal processing approach to fair surface design.
\newblock In \emph{Proceedings of the 22nd Annual Conference on Computer Graphics and Interactive Techniques} (New York, NY, USA, 1995), SIGGRAPH '95, Association for Computing Machinery, p.~351–358.
\newblock URL: \url{https://doi.org/10.1145/218380.218473}, \href {https://doi.org/10.1145/218380.218473} {\path{doi:10.1145/218380.218473}}.

\bibitem[WMKG07]{noFreeLunch:07}
\textsc{Wardetzky M., Mathur S., Kälberer F., Grinspun E.}:
\newblock Discrete {L}aplace operators: No free lunch.
\newblock vol.~07, pp.~33--37.
\newblock URL: \url{https://doi.org/10.2312/SGP/SGP07/033-037}.

\bibitem[WMVB23]{walker2023}
\textsc{Walker T., Mariotti O., Vaxman A., Bilen H.}:
\newblock Explicit neural surfaces: Learning continuous geometry with deformation fields, 2023.
\newblock \href {http://arxiv.org/abs/2306.02956} {\path{arXiv:2306.02956}}.

\bibitem[WZL{\etalchar{*}}18]{pixel2mesh:18}
\textsc{Wang N., Zhang Y., Li Z., Fu Y., Liu W., Jiang Y.}:
\newblock Pixel2mesh: Generating 3{D} mesh models from single {RGB} images.
\newblock \emph{CoRR abs/1804.01654} (2018).
\newblock URL: \url{http://arxiv.org/abs/1804.01654}.

\bibitem[XZ03]{xu2003eulerian}
\textsc{Xu J., Zhao H.}:
\newblock An {E}ulerian formulation for solving partial differential equations along a moving interface.
\newblock \emph{Journal of Scientific Computing 19} (2003), 573--594.
\newblock \href {https://doi.org/10.1023/A:1025336916176} {\path{doi:10.1023/A:1025336916176}}.

\bibitem[YBHK21]{yang2021geometry}
\textsc{Yang G., Belongie S., Hariharan B., Koltun V.}:
\newblock Geometry processing with neural fields.
\newblock In \emph{Thirty-Fifth Conference on Neural Information Processing Systems} (2021).
\newblock URL: \url{https://dl.acm.org/doi/10.5555/3540261.3541983}.

\end{thebibliography}


\end{document}